# INATTENTIONAL BLINDNESS WITH AUGMENTED REALITY HUDS: AN ON-ROAD STUDY

**Nayara de Oliveira Faria & Joseph L. Gabbard**
VIRGINIA TECH

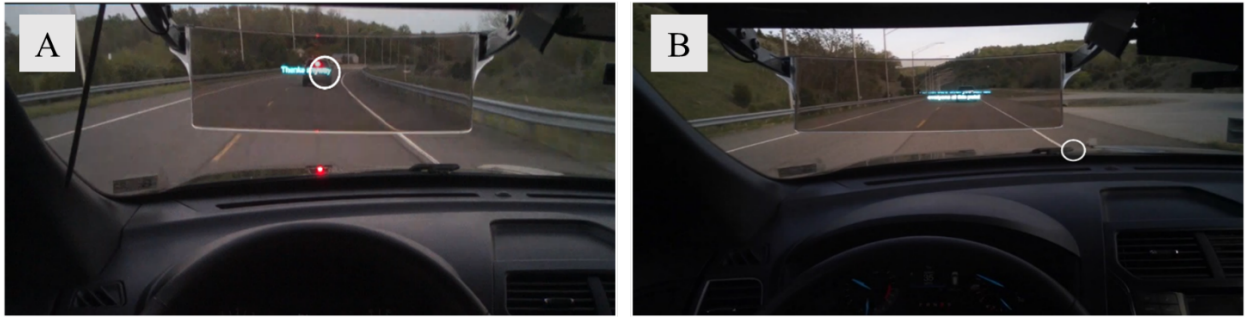

*Figure 0-1. Driver engaged in a secondary augmented reality task of reading a text message using a head-up display in a real-world driving scenario on actual roadway. A: **Inattentional Blindness** - the driver's gaze is directed towards the lead vehicle's brake light, which is illuminated, but the driver fails to notice it. B: **Attention Capture** - the driver's response to a stimulus on the periphery of the road is delayed due to the presence of the AR secondary task.*

## 1. INTRODUCTION

As the integration of augmented reality (AR) technology in head-up displays (HUDs) becomes more prevalent in vehicles, it is crucial to understand how to *design* and *evaluate* AR interfaces to ensure safety. With new AR displays capable of rendering images with larger field of views and at varying depths, the visual and cognitive separation between graphical and real-world visual stimuli will be increasingly more difficult to quantify as will drivers' ability to efficiently allocate visual attention between the two sets of stimuli.

In comparison to traditional in-vehicle interfaces, AR interfaces pose a challenge as they blur the line between synthetic and real-world scenes, lacking the visual separation between the interface and driving scene. AR interfaces, especially world-relative, exist within the line of sight needed to perform the primary visual driving task; and moreover, these interfaces may be present independent of whether or not drivers should be attending them. In short, we know that the blending of synthetic and real-world scenes creates both opportunities and safety concerns for AR use in safety critical situations.

One of the main safety challenges with AR in driving is the potential for its graphics to demand or capture drivers' attention and subsequently distract drivers from perceiving and responding to important events and objects in a timely manner. This momentary inability is essentially artificially induced *inattentional blindness.* Inattentional blindness (IB) often described as "looking without seeing" [1] – occurs when human gazes are directed in the general vicinity of some stimuli, yet only a small amount of information is genuinely perceived [2]. As such, while performing secondary AR tasks, drivers may overlook important real-world events, such as a shift in traffic light color [3], a pedestrian crossing or a lead car's brake [4]–[7]. Early studies revealed that the utilization of AR HUDs in aircrafts increased the likelihood of inattentional blindness as pilots were found to overlook other aircrafts on runways, despite looking in that direction [8]. However, these findings cannot be extrapolated directly to surface transportation as the driving environment and usage of these displays are vastly different. For example, aviation HUDs typically show runway or horizon outlines, at much greater distances whereas AR in vehicles is expected to be more versatile, showing virtual information as our cellphones do, in the near-to medium field and in fast-paced, life-threatening situations. Therefore, in a design space that affords a visual integration of AR graphics and the real-world dynamic environment, inattentional blindness **must** be considered in the design process of AR interfaces as it can offset the benefits of using AR while driving by introducing new risks.

Designing AR interfaces that provide constant access to a vast amount of information in dynamic environments while minimizing the risk of inattentional blindness remains a challenge. To address this issue, a deeper understanding of both intrinsic AR interface factors and environmental factors contributing to inattentional blindness is necessary. By gaining this knowledge, we can create intelligent and adaptive AR systems that effectively reduce the likelihood of inattentional blindness and better cater to human cognitive limitations when using the technology.

## 1.1 OBJECTIVES AND CONTRIBUTIONS

In this study, we present a user study that serves as a crucial first step in gaining insight into inattentional blindness while using AR in surface transportation, where understanding is currently limited. Our primary goal is to investigate how the visual demand of AR tasks influences drivers' ability to detect stimuli, and whether the nature of the stimuli itself plays a role in this effect. To address these questions, we designed an on-road user study aimed at producing a more realistic and ecologically valid understanding of the phenomenon.

In this study, participants were asked to simultaneously read AR text messages, identify stimuli on the forward roadway, and drive a real vehicle. The key contributions of this research include:

- **An ecologically valid detection methodology** that moves beyond conventional surface transportation research methods. Our approach assesses inattentional blindness by evaluating drivers' ability to detect critical stimuli on the road while using an AR display.
- **An on-road AR driving study** conducted in a real-world environment, in contrast to simulator or virtual reality setups. By studying drivers in actual driving conditions, we aim

- to gain richer insights into the mechanisms of visual attention and how they relate to inattentional blindness when interacting with AR systems.
- **The foundation for a safety-centric framework** that maps the relationship between key factors contributing to inattentional blindness, with a focus on drivers' visual behavior and their capacity to perceive relevant events and objects on the road. This framework is intended to inform the development of improved safety assessment tools—positioning inattentional blindness as a critical performance metric for evaluating the safety of AR technologies in transportation contexts.

## 2. RELATED WORK

### 2.1 INATTENTIONAL BLINDNESS & THE COGNITIVE PSYCHOLOGY PERSPECTIVE

Several studies from the 1970s and 1980s have shown that conscious perception appears to require attention [9]–[12], and as such, no perceptual object can exist preattentively [13]. In this way, when attention is diverted to another continuous task or object, people frequently fail to notice an unexpected event, even if it is located at fixation; a phenomenon later defined as *"inattentional blindness"* [2]. Accordingly, although there is a relationship between where people fixate and where they attend to [14]; the phenomenon of inattentional blindness illustrates that attention and fixation can be separated from each other [15].

These studies of inattentional blindness were designed to investigate how people process and select features and objects, a phenomenon defined as "selective looking" at the time. In these studies, participants were required to attend to a continuous monitoring task focusing on a specific part of a scene while ignoring the other parts. An unexpected event occurs at some time during the monitoring activity, but the majority of participants do not report seeing it, despite the fact that it is readily apparent to other people who are not engaged in the main task [9]–[12].The experimenters used simple visual displays at precisely controlled timing conditions; an approach that was developed to be a visual mimic of dichotic-listening studies carried out in the 1950s and 1960s [16]–[18]. Although this computer-based approach was quite successful in demonstrating the phenomenon of selective-looking, most studies used static events rather the dynamic ones. To address this concern, Niesser conducted a series of studies in the 1970s.

In the first study [11], participants were asked to look at two superimposed video screens in which simultaneous but different events were taking place. One event was a "handgame" in which one player tries to slap his opponent hand; the other was a "ballgame" in which three players pass a basketball to each other while moving around the room in irregular patterns. Participants were required to monitor either events and press a button when a relevant action occurred, either a slap or a ball passing. The results of this study are largely consistent with the findings of selective listening: people can easily follow one visual event while ignoring the overlapped one as they are to listen to one of the two simultaneous conversations. The same phenomenon can be observed even if participants are prohibited from moving their eyes (see [10]). In this way, even when no unique filtering mechanism has been created via experience and whether eye movement is involved; Niesser & Becklen [11], emphasize that perception is selective.

In one subsequent study, Niesser investigated whether selective looking depends on the similarity of the simultaneous events taking place (see Becklen, Neisser, and Littman, discussed in Neisser, 1979). In this new version of the work, they introduced two salient stimulus features (presence/absence of sunglasses and black/white shirts) by which the superimposed teams could be distinguished. At this time, the same ballgame was being played by both teams, and participants were required to press a button when the target team made a ball pass. Although the modification of the study made ignored and target events more similar, participants were able to successfully follow one team while ignoring the other.

In the most famous variation of the study [19] , the phenomenon of inattentional blindness was first reported in the literature. In this experiment, participants watched a ballgame video in which two teams wearing different shirt colors (black and white; highly salient attribute) were playing. The video of each team was separately recorded and then electronically superimposed to make a "stimulus video" containing both teams (six people and two balls).  If two players occupied the same space on the screen, they appeared to pass through each other in a "ghostly" way. Participants were instructed to ignore the white shirt players and to push a key anytime the black shirt players successfully passed the ball.  Approximately halfway through the video, a woman carrying an umbrella passed by the scene for approximately four seconds. Participants were so fixated on identifying passes that they frequently missed the "umbrella woman".  Further, this experiment evidenced that noticing unexpected events is highly correlated with task perceived difficulty; the easier the task becomes or seems to become, the more likely people are to notice other things. This phenomenon could be attributed to a drop in attentional capacity sufficient to sustain no more than performance on the primary task, making the unexpected event less likely to be detected [20].

Neisser [19] presented another variation of the study in which the umbrella woman wore the same color shirt as either the attended or unattended players. The manipulation of this "feature-similarity" seemed to have no influence on the rate at which participants perceived the unexpected event. In addition, when the experimenter switched the umbrella woman for a small boy drinking a can of soda, fewer participants noticed him. Further, when the umbrella woman performed a little dance instead of just passing by the scene, more subjects noticed her. Altogether, these findings suggested that although the similarity to attended stimuli appears to make little difference in the rate of noticing; the properties of an unattended stimulus can capture attention.

In spite of Niesser's work addressing the concern of investigating selective looking in more dynamic scenes instead of static ones; perhaps the most important unanswered question and criticism of his early work is whether inattentional blindness was influenced by the superimposition of videos. Specifically, the umbrella woman and the players were less visible than they would be in the real world without the superimposition of events.

To address this concern, in the late 1990s, Christopher Chabris and Daniel Simons [21] replicated Neisser's original work. The authors found that when both white and black shirt actors were partially transparent and occupied the same area, most participants missed the umbrella woman. Further, Chabris and Simons introduced a new version of the stimulus video in which both teams were filmed using a single camera; in a way that players were always visible and never "ghostly" passed through each other. In this new version of the study, a woman in a gorilla costume walks into the scene, stops to face the camera, pounds her chest, and then walks out to the opposite side after nine seconds on screen. Although the gorilla passed across the participants' line of view, over half of them failed to notice it while counting passes by the white team. Further, this experiment revealed four other important findings: 1) the level of inattentional blindness varies according to the difficulty level of the primary monitoring task; 2) participants are more likely to perceive an unexpected event if it is visually similar to the events their attention is focused to. By focusing on shared cues between primary task and unexpected event, attentional resources are freed up, allowing the unexpected event to be noticed more easily [20]; 3) although people frequently miss objects appearing in live-action opaque displays, inattentional blindness occurs in a higher frequency in transparent displays in which actors move through each other; 4) the proximity of the unattended event and the location where participants are attending to does not seem to influence perception, suggesting that people attend to objects and events rather than spatial positions.

Pappas et al. [22] extended the "Invisible Gorilla" study [21], applying eye-tracking technology to investigate the influence of proximity and similarity of stimulus to the unattended object. Overall, the authors found that participants counting passes made by black shirt players had a higher rate of noticing, indicating that the similarity of stimulus to the unattended event improves inattentional blindness. Also, it was found that although participants were fixing at the stimulus of interest, many of them failed at noticing it. In contrast, other participants who noticed the gorilla in the video had no fixation towards it, indicating that perception was acquired using peripheral vision.

## 2.2 INATTENTIONAL BLINDNESS IN DRIVING

Although inattentional blindness in driving has the same theoretical foundation as cognitive psychology research, there are significant inherent distinctions between them, making it difficult to explain why and how we attend to some items and not others in real-world settings. Even in its simplest form, a driving environment is a more complex activity than full attention tasks performed in cognitive psychology studies (i.e., counting ball passes, counting hand slaps, counting the number of letters in a screen). Also, drivers are constantly required to perceive and respond to a variety of dynamic visual events while engaging in other nondriving-related tasks. Understanding inattentional blindness in cognitive psychology studies has helped in elucidating the influence of unexpected stimulus features on perception and demonstrate important real-world implications [23]; nevertheless, this approach is inadequate for investigating the phenomena in more realistic and dynamic scenarios.

In driving research, inattentional blindness is often termed as a *"looked-but-failed-to-see"* (LBFTS) [24] driving error. The term was first introduced by Treat [25], long before AR UIs were integrated into the driving scene, to refer to drivers' inability to detect another roadway object despite looking at it. Brown [26] analyzed UK driving accident data and found that LBFTS crashes were the third most frequent type of accident; and the largest contributing factor when intoxication, night driving, and illness were removed from the dataset. It is believed that LBFTS errors usually occur when drivers are either novices or older [27], [28] at intersections [28], in dual-tasks [28], or in attention-demanding situations [28].

Another important factor that influences driver's ability to perceive a potential target is its expectancy, which is formed by both long-term driving experience as well as the immediate experience from the past few minutes of driving [29]. In this way, if a driver is scanning the roadway for cars or larger vehicles, since "there is a higher probability that any potential conflict is going to be with these types of road elements" [29, p. 193]; the driver may look at a cyclist or motorcyclist and then drive directly into them. However, it is thought that the greater the number of bicycles and motorcyclists, the less this expectation problem will exist [29]. Other studies have also shown that low expectancy levels result in higher levels of LBFTS errors, measured in terms of detection distance to an object [30], [31]. Significantly, a comprehensive meta-analysis conducted by Green [32] pinpointed expectancy as the predominant factor influencing braking reaction times in response to critical events while driving, revealing that completely unexpected events take twice as long to be detected as anticipated ones.

Moreover, it is proposed that the intrinsic semantic content of stimuli plays a crucial role in how attention is captured in inattentional blindness situations. This means that a child standing by the road is more likely to attract attention and be detected by a driver than a less significant object like a garbage bin in the same location [33]. As Pammer et al. elaborate [33, p. 783] " *if a contextually appropriate object is placed in a visual scene, and the type of object is varied only in regards to its importance to the environment, then we would hypothesize that a contextually relevant coding mechanism would scan the scene, selecting those objects that are most important and filtering out those that are less important*".

Finally, another important factor influencing inattentional blindness in driving research is cognitive load. When performing cognitively demanding tasks, drivers show slower reaction time to a lead vehicle brake [4]–[7], higher levels of missed critical stimuli relevant to the driving task [34], and diminished sensitivity to events in the periphery [35]–[38].

### 2.3 INATTENTIONAL BLINDNESS IN HEAD-UP DISPLAYS

Early evidence for inattentional blindness while using AR was found in a study conducted by NASA-Ames that delivered visual information via head-up displays [8]. The study found that response time to an unexpected event (an airplane taxing into the runway) was longer when using a HUD compared to a traditional instrument head-down display. Furthermore, two out of eight pilots did not detect the aircraft at all. Despite the lack of statistical analysis due to limited experimental power, the authors believed that pilots using the HUD were less likely to identify unexpected events in the far domain because their attention was improperly focused on the HUD symbology. The format of the instrumentation in the head-up and head-down instruments, however, made drawing conclusive results of the effects of superimposition difficult. To address this concern, Wickens and Long [39] employed higher statistical power and equated the instrumentation format across the head-up and head-down positions to replicate this study utilizing a high-fidelity flying simulator. They found that pilots were slower to respond to a low likelihood unexpected aircraft on the runaway when the symbology was in the head-up location as compared to the identical symbology presented in the head-down location, supporting Fischer et al. [8] findings.

Other studies have also demonstrated the same trend of longer latency to detect unexpected events when using HUDs [40]–[42]. Further, Wickens et al. [43] highlight three key factors related to inattentional blindness when using HUDs: 1) the phenomenon only seem to occur in the face of truly unexpected events, such as a sudden runway incursion that catches a pilot off guard; 2) the likelihood of experiencing inattentional blindness seems to be reduced with the use of conformal graphics [39], [44], as this symbology provides a direct linkage or overlay between near and far domains ("*scene linking*" that facilitates divided attention [45], [46]; 3) the use of the conformal tunnel graphics during flight appears to increase inattentional blindness [44], as this type of symbology presents a highly compelling, self-centric guidance for attention that lacks a corresponding visual representation in the distant domain, thereby attracting attention in a manner akin to non-conformal symbols.

## 2.4 INATTENTIONAL BLINDNESS IN AUTOMOTIVE AUGMENNTED REALITY HEAD-UP DISPLAYS

While the understanding of inattentional blindness with head-up displays in the aviation industry provides valuable insights into the intrinsic aspects of the display, the findings may not be directly applicable to the use of this technology in the automotive industry. This is due to several key differences between the two contexts. Firstly, the external environment in aviation is typically composed of low salience clouds and open sky, while in driving, drivers encounter fast-moving, highly salient and often life-threatening objects and events. Secondly, AR delivered via head-up displays are often used as a primary source of information by pilots, whereas in automotive applications, they are more likely to be used as a secondary source of information by drivers. Lastly, in aviation, HUDs typically use conformal symbology, such as a runway outline, while in driving, information is presented in a non-conformal manner, such as text or 2D graphics. These discrepancies highlight the need for further research specifically focused on inattentional blindness in the automotive context.

Drivers receive AR information from heads-up displays without diverting their gaze from the road, making use of the proximity-compatibility-principle [47], a crucial aspect in in-vehicle display design [48]. However, even if the AR graphics are presented at the same focal depth as real-world references, there is a cognitive cost to switching between the two [49], potentially leading to inattentional blindness. The size and prominence of the display imagery can also hinder drivers' perception of the far domain [50] , but the exact impact of display imagery saliency on drivers' detection capabilities is unknown. In the same way, although the advantages AR information delivered via HUDs are moderated by two key display features: conformality and information integration [50]; the relationship between these features and inattentional blindness while driving remains to be explored.

Previous research has shown that AR HUDs significantly affect driver visual behavior compared to traditional in-car displays. The increased visual attention towards the AR graphics [51]–[53], as indicated by longer glances, may lead to drivers focusing on the AR graphics without fully processing roadway information [54], presenting a potential danger in driving situations. To address this issue, much effort has been put into developing AR interfaces that can effectively redirect driver attention to potential hazards in the environment. These interfaces, such as forward collision warnings [55], blind spot warnings, and pedestrian collision warnings systems [56], [57] use AR cues to alert drivers of potential undetected stimuli and prompt appropriate responses. Indeed, in two recent studies exploring inattentional blindness with augmented reality head-up displays [58], [59] researchers discovered that the utilization of conformal boxes to accentuate potential road hazards (such as pedestrians or a leading vehicle braking) effectively diminishes the probability of drivers experiencing inattentional blindness.

In the near future, AR in vehicles is expected to display information in a similar manner to cellphones, making it imperative to understand how to design tasks to prevent inattentional blindness. For instance, Smith et al. [60] conducted a study to examine the effects of AR graphics location on drivers' ability to detect unexpected brake events from a lead car. They found that when the display was located in the middle of the windshield, drivers were more likely to collide with the lead car. They suggest that adding AR information into or over other tasks may cause drivers to miss critical cues in the environment because their attention is split between tasks [54], [61] However, previous research [62], [63] has shown that AR graphics located at lower eccentricities (such as the middle location studied by Smith et al. [60]) improve lateral driver performance which is the main metric of safe driving performance. This highlights the need for further investigation into the relationship between AR and drivers' ability to perceive significant events and objects while utilizing the technology for everyday tasks.

## 3. USER STUDY

### 3.1 RESEARCH QUESTIONS

Our on-road user study aimed to investigate intrinsic AR interface factors contributing to inattentional blindness and their interaction with external environmental factors. We evaluated drivers' ability to detect stimuli on the road while using AR (e.g., likelihood of inattentional blindness) and posed two key research questions: RQ1 explored intrinsic AR interface factors, and RQ2 examined the influence of external environmental factors.

- **RQ1:** How does AR task visual demand influence the detection likelihood of stimulus on the roadway?
    - **H1:** The level of inattentional blindness will increase as the visual demand of the AR task increases.

We designed three levels of AR task complexity to induce varying degrees of visual demands: low (e.g., one-line text), medium (e.g., two-line text), and high (e.g., three-line text). In this way, visual demand is related to the amount of information being presented to drivers at any single point in time. This approach allows us to study whether or not presenting the same information in smaller chunks, will decrease the likelihood of inattentional blindness – a key insight for designing future AR UIs.

To design the AR task for this study, we carried out a preliminary user study employing a driving simulator. The measurement of visual demand was achieved through a dual-task paradigm, comparing driving performance across various AR task levels against a baseline condition. Our primary objective was to ensure that visual demands at each level could be clearly differentiated, while also preventing the induction of hazardous driving behaviors as this study would be conducted in a real-life driving scenario.

- **RQ2:** How is the likelihood of inattentional blindness (when using AR) influenced by the type of real-world stimulus to be detected?

In examining this research question we were also interested in understanding the impact of stimuli location and type on drivers' ability to perceive them while using AR. Specifically, we designed the set of stimuli to better understand if stimuli in drivers' central field of view are more likely to be missed as compared those in the drivers' peripheral field of view, and if some stimuli are more prone to being missed than others (e.g., due to their perceived value). Generally speaking, we still have much to learn on how well we can leverage AR graphics to guide visual attention to important real-word hazards, and to what extent the visual onset of AR graphics draws attention away from the driving. By understanding stimulus locations and types more or less prone to be missed, the AR systems can be adapted to suit environmental demands and human cognitive limitations. For this research question, we posed two main hypotheses:

- **H2A**: Stimuli in the peripheral field of view are more likely to be missed by drivers compared to stimuli in the central field of view.
- **H2B**: Stimuli with lower perceived value are more likely to be missed by drivers compared to stimuli with higher perceived value.

### 3.2 PARTICIPANTS

Our user study included twenty-four participants, evenly divided by gender with 12 males and 12 females, ranging in age from 19 to 44 years ($\mu = 23.08$ years, $\sigma = 5.92$ years). All participants held a valid US driver's license and had corrected visual acuity of at least 20/40 (6/12) as determined by the Snellen acuity test. The study was reviewed and approved by the Virginia Tech's Institutional Review Board (IRB # 22-294) and all participants were compensated $30 per hour for their time.

### 3.3 EXPERIMENTAL DESIGN

We conducted a within-subjects, three-factor repeated measure experiment in which participants engaged in a tertiary-task study. This study was conducted on a controlled roadway, termed the Smart Road, under optimal weather conditions of daylight and clear skies (no rain, snow, or fog). The *Smart Road* is a 2.2 mile-long (3.5km) controlled-access research facility that adheres to U.S. highway specifications, with wide lanes, shoulders, and clear white markings the road's exterior margins.

The *primary driving task* was a car-following task on a two-lane highway with no other vehicle present except for a lead vehicle, which travelled at a constant speed of 35mph at a distance of 150 meters. We decided not to include any additional road traffic, as past research has shown that high environmental driving demands can greatly affect drivers' ability to detect targets [64], [65]. To minimize this potential confound, we chose a low-demand driving environment which is consistent with NHTSA's current specifications for assessing safety of in-vehicle display designs.

As participants carried out the primary driving task, they simultaneously engaged in a *secondary AR task* and, a *tertiary detection task.* Our study manipulated the visuals associated with the secondary AR task, the type of detection stimuli present in the tertiary task, and used gender as a blocking variable.

- **Gender (2 levels, block):** we recruited twenty-four participants who were evenly distributed in terms of gender self-identification.
- **AR task visual demand (4 levels, within-subjects):** our secondary task involved an AR text message task, which varied in three levels of visual demands. These levels ranged from low (e.g., short, one-line text) to high (e.g., long, three-lines text). We also included a baseline condition *("baseline"),* where participants did not engage in any AR secondary visual task.
- **Detection task stimulus (4 levels, within-subjects):** We used three types of detection stimulus: the third brake light of the lead vehicle (*"light"*), a child mannequin *("child"),* and a wooden target *("target")*. These stimuli were placed in either the central or peripheral field of view of participants, with the aim of assessing *central* and *peripheral* detection task performance. We also included a baseline condition ("*baseline"),* where participants did not engage in detecting any stimulus on the road.

We employed a nested counterbalanced experimental design that utilized 2x4 and 3x4 configurations with two repetitions each. This design enabled us to effectively study the impact of AR visual demand and detection stimuli on both the central and peripheral field of view (as shown in Table 0:1). The experimental design generated a total of 40 possible event combinations, each representing a unique pairing of AR, baseline, and detection tasks. From this total, 24 events represented a detection task event combined with AR tasks or baseline conditions. To control for potential order effects and enhance the internal validity of the study, we utilized a Latin Square design to randomly order the 40 events, resulting in six possible event orders that were randomized across participants.

*Table 0:1. The experimental design we used in this study. The central detection has a 2x4 nested counterbalanced design with two repetitions and a total of 16 events per participant, from which 8 events represented a central detection task event. The peripheral detection task has 3 x4 nested counterbalanced design with two repetitions and a total of 24 events per participant, from which 16 events represented a peripheral detection task event.*

| CENTRAL DETECTION TASK (2x4 DESIGN) | | PERIPHERAL DETECTION TASK (3X4 DESIGN) | |
|---|---|---|---|
| **Detection Task Stimulus** | **AR Task Visual Demand** | **Detection Task Stimulus** | **AR Task Visual Demand** |
| Light | Low (1 line) | Target | Low (1 line) |
| Baseline | Medium (2 lines) | Child | Medium (2 lines) |
| | High (3 lines) | Baseline | High (3 lines) |
| | Baseline | | Baseline |

## 3.4 THE AR TASK

Participants performed a *secondary AR task* concomitant with the primary driving task. We chose the *text message task* as the secondary task based on three main points: 1) it is a task of low priority, 2) it is a task that is not integral to a critical driving situation, and 3) the task is ecologically valid given the increasing integration of AR into vehicles. While for this study, the text is presented in 2D, we posit that not only is this an ecologically valid task, but that: (1) AR interfaces include a combination of 2D text elements (e.g., labels, notifications, messages) and 3D objects, and, (2) our approach and findings are applicable to 3D objects presented on the front roadway, and (3), our approach and findings are applicable to AR information presented via AR HUD or AR head-worn display. In this task, participants were required to read aloud an incoming English text message, word by word, to complete the task. The text stimuli consist of statements, not questions, as we are only interested in investigating secondary task visual demands and not in creating additional cognitive load associated with long-term memory retrieval. Text messages were retrieved from the Enron Mobile Email Dataset (see [66] for details on the dataset). We defined three levels of task visual demands: low (e.g., one-line text), medium (e.g., two-line text), and high (e.g., three-line text).

## 3.5 THE DETECTION TASK

Before delving into the specifics of the detection task approach employed in this study, it is crucial to first establish a foundational understanding of the *Detection Response Task* (DRT) – a widely utilized method in the field of driving research. By providing a comprehensive overview of this method, we aim to provide a solid foundation that will enable us to effectively explain and justify the experimental design choices we made for this study.

### 3.5.1 THE DETECTION RESPONSE TASK

DRT is a low-cost method for evaluating the effects of secondary tasks' cognitive load demands on driver attentional resources [67], [68]. In this dual-paradigm method, the driver is required to perform a primary driving task and to detect and respond as quickly as possible to frequent and random stimuli (secondary task). The premise of the DRT is that cognitive interference arises when several tasks impose simultaneous demands on cognitive control, and not enough cognitive resources are available to support those tasks in parallel [68]. In this way, the ISO standard, ISO/DIS 17488 [68], was created to assist automobile and aftermarket in-vehicle display manufacturers in evaluating the cognitive workload of an interface using DRT. According to the specifications of the standard, the DRT stimuli, which might be visual (LEDs), tactile (vibration motor), or audio (blip), must emerge at random time intervals every 3–5 seconds, and participants must respond to them by pressing a micro-switch button. The secondary task cognitive load is then evaluated based on participants' response times and hit rates. If the goal is to evaluate visual demands related to glances toward a visual display, then the recommendation is to use the remote DRT method. This method is further divided into *peripheral detection task* (PDT), using one or more visual probes in the peripheral visual field; and visual detection task (VDT), using one forward visual probe in the central visual field. In this study, the visual detection task is referred to as *central detection task* (CDT).

### 3.5.2 THE PERIPHERAL DETECTION TASK (PDT)

Peripheral detection tasks were created to assess detection performance in the visual periphery, and one of the task's basic assumptions is that high workload leads to *tunnel vision* [69], [70], which is the reduction of visual sensitivity on the visual periphery [71]. This method has been widely employed in in-vehicle display safety testing because it is thought that the capacity to respond to basic visual stimuli is similar to reacting to the quick appearance of objects and events on the actual roadway [72]. In traditional PDT studies, participants are required to respond to artificial red lights randomly appearing to the left or right of the road scene, usually at a horizontal angle of 11° to 23°and a vertical angle of 2° to 4°above the horizon. A pedal response [73]–[79] or pressing a button on the driver's index finger [69], [72], [80], [81] are the two main methods of responding to PDT stimuli. Also, the driving task is a simulated or surrogate driving task [69], [72], [81], [73]–[80] where participants watch a pre-recorded road scene while performing the stimuli detection task.

This method has been criticized because it demands drivers to respond to stimuli that are shown on a regular basis, and drivers have some knowledge of their spatial and temporal location; a situation that is not ecologically valid with the real road.  Also, the visual detection task stimuli are placed in a position relative either to the observer or the observer's moving eye gaze (as opposed to in the road scene).  In the driving context, research suggests that drivers use saccades to perceive relevant information from the driving scene [82], [83].  Therefore, we decided to place stimuli directly on the periphery drive scene so that we can examine how AR affect drivers' ability to detect important elements outside the display's field of view. With this approach, the apparent stimuli eccentricities and apparent size will increase as drivers move along, as opposed to most previous studies where target stimuli remained at a fixed location and with a fixed apparent size.

We used two types of stimuli for the peripheral detection task: Adrian's [84] small target (referred to as "target") and a mannequin of a child (referred to as "child"). Adrian's [84] small target visibility model is a well-established approach that suggests that drivers' ability to detect a small squared target on the side of the road is a direct quality measure for assessing the visibility of a particular roadway lighting system. Therefore, during daylight hours, the ability to detect small square targets can directly measure a driver's visual performance and decreases in performance can be attributed to the use of the AR. We decided to use both types of stimuli for the periphery because previous research suggests that the intrinsic semantic content of the stimuli [33] captures attention differently. Therefore, we expect that when using AR, a child standing on the side of the road would be more likely to be detected than an unimportant small target placed in the same location.

We utilized gray, 18-cm (7-inch) square targets with a 9-cm (3-1/2-inch) square protrusion on one side (Figure 0-2A) that were positioned 0.3 meters outside the right shoulder of the road, as done in earlier work [30], [85]–[89].  The child-sized mannequin measured 1.2 meters in height, which were outfitted in gray-colored scrubs as shown in Figure 0-2B. These mannequins were placed at the same location as the targets, 0.3 meters outside the right shoulder of the road.

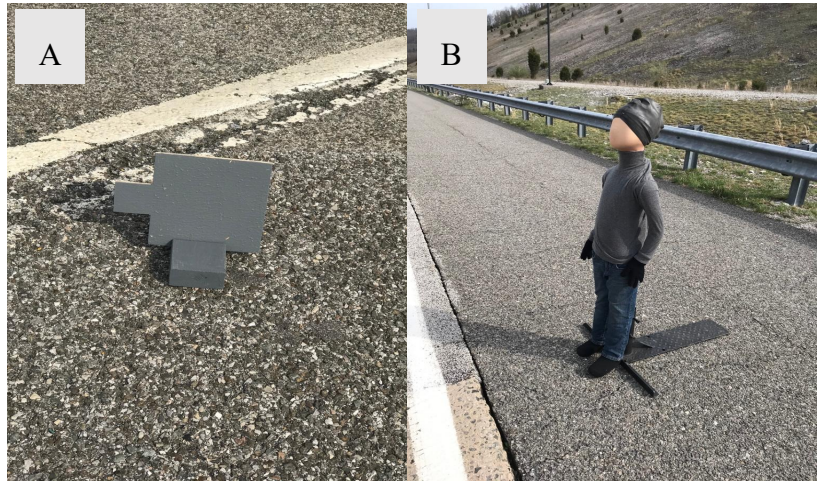

*Figure 0-2. Peripheral detection targets used in this study: A) Adrian's small target (referred to as "target") and, B) Child-sized mannequin (referred to as "child").*

### 3.5.3   THE CENTRAL DETECTION TASK (CDT)

Central detection tasks were created to assess detection performance in the central field of view. This method mainly differs from the original peripheral detection task in that it only involves a single stimulus located in the driver's central field of view rather than the periphery [68]. Also, the stimulus used for the CDT has higher LED intensity compared to the ones used in the PDT. This change was made to increase stimulus saliency while limiting the impact of ambient lighting conditions and reducing the requirement to glance directly at the stimulus to detect it [71]. Another approach for CDT is to use the lead vehicle brake light as the stimulus [4]–[7]  In this study, we used the third brake light of the lead vehicle as the detection stimulus in the central field of view of participants.

### 3.5.4   OUR DETECTION APPROACH

Our proposed detection task approach differs from traditional methods in several ways. Firstly, the intrinsic semantic meaning of our stimuli, being a more natural component of the driving task, sets them apart from traditional methods. We aimed to create salient stimuli that are relevant to driving but did not interfere with the driving task, as it might otherwise reorient a driver's attention that was directed somewhere else at the time. Additionally, the regular and less frequent presentation of our stimuli makes them less predictable for drivers. Furthermore, detecting a lead vehicle brake light or a child on the side of the road are considered less expected events on the expectancy continuum [71], as opposed to signal detection tasks as prescribed by the ISO standard ISO/DIS 17488 [68]. This suggests that our proposed detection task have a good deal of ecologically validity. We believe that the results of this study will have immediate implications for driving safety when using AR, as it will indicate the extent to which drivers are delayed or fail to detect important information on the environment.

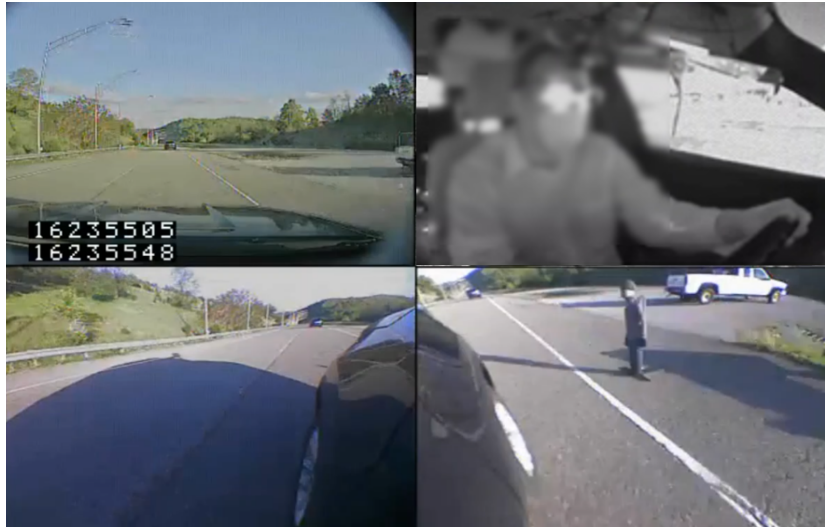

*Figure 0-3. Data acquisition system used in this work. The bottom right video was used to evaluate when drivers passed stimuli on the periphery of the road.*

### 3.6 EQUIPMENT AND APPARATUS

Participants in the study drove a 2017 Ford Explorer that was outfitted with a data acquisition system (DAS) to collect kinematic data from the vehicle's Controller Area Network (CAN) system. This included information on the vehicle's speed, GPS coordinates, steering pitch, and four video images from different perspectives (driver's face, forward roadway, left side of the roadway, and right side of the roadway, as shown in Figure 0-3). The DAS also recorded manual button presses and other inputs from the in-vehicle experimenter, which were used to verify when a participant detected a stimulus on the roadway. Furthermore, a confederate driver operated a similar 2017 Ford Explorer, serving as a lead vehicle for the participants to follow throughout the study.

We equipped the experimental vehicle with a Pioneer Cyber Navi head-up display with conformal augmented reality graphics capabilities, as shown in Figure 0-4. The AR HUD had a display area of 780x260 pixels, a field of view of 15 degrees, an accommodative demand of approximately 3m, with the virtual image positioned approximately 3m away from the eyepoint. Participants were also fitted with Tobii Pro Glasses 2 eye-tracking systems, which were equipped with audio and video recording capabilities to track gaze behavior throughout the study.

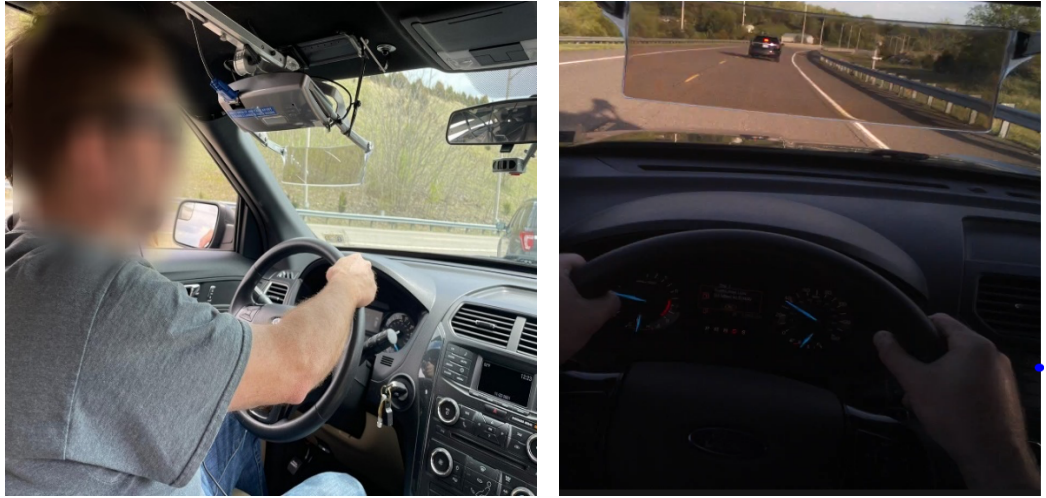

*Figure 0-4. Driver equipped with Tobii Pro eye-tracking glasses, executing the primary driving task of following a lead vehicle via and using an AR Heads-Up Display.*

## 3.7 PROCEDURE

Before starting the study, participants completed an intake session where they were provided informed written consent and underwent a visual acuity test using the Snellen method to ensure they met the requirements for the study. Additionally, we asked them to complete a series of questionnaires that included demographic information, driving risk [90], and susceptibility to driving distractions [91]. Once the intake session was completed, we escorted the participant to the experimental vehicle parked outside. The experimenter demonstrated the proper adjustments for the seat and steering wheel and instructed the participant to make any necessary adjustments before buckling their seatbelt.

The in-vehicle experimenters situated themselves in the back seat of the vehicle, setting up the necessary data collection equipment. Once the Driver Assistance System was prepared, we instructed the participant to follow the lead car on the Smart Road, maintaining a speed limit of 35 mi/h (56 km/h) throughout the study. To acclimate the participants to the road and detection objects, the first lap served as a practice lap, allowing them to familiarize themselves with the driving task at hand.

The experimental phase began once the participants indicated they were ready. We instructed participants to park their vehicle 150m behind the lead vehicle. Then we fitted them with Tobii eye-tracking glasses and performed a HUD calibration process to ensure the text messages were displayed in the same location for all participants. Next, we introduced the head-up display and provided clear instructions for the three tasks to be completed: *primary driving task, secondary AR task,* and *tertiary detection task.*

During each lap, participants were presented with AR secondary tasks of varying levels of visual demands. For each text message event, we instructed participants to read as many text messages as possible during the event. During an event, once participants finished reading a given text message, we pressed a button on the laptop to record the duration of that specific text reading task and then immediately presented the next text message associated with the event. In this way, particiants' chose the pace (and in term the number of messages) they read during each event.

In addition to the AR tasks, participants were also expected to attend to central and peripheral detection tasks, to which they were instructed to respond appropriately. Specifically, for the peripheral detection task, we prompt participants to verbally state "pedestrian," "kid," or "child" (whichever was easiest for them to remember) upon sighting one of the child-sized mannequins and indicate the side of the road the mannequin was facing (left or right). We recorded the identification by pressing a handheld button and noting the side of the mannequin. The same instructions were given for the detection of the target, where participants were prompted to say "target, left" or "target, right". For the central detection task, the lead car turned on its third brake light for approximately 2 seconds, we instructed participants to say "light" as soon as they saw it. The presentation of detection tasks and AR tasks were counterbalanced to reduce potential order-related confounding. After each lap, participants were given a questionnaire (adapted from Bhagavathula et al. [92]) to evaluate how they prioritized the tasks, and their situation awareness during that experimental lap.

We asked participants to complete a post-trial questionnaire after finishing the experimental session where they provided feedback on their perceptions of the head-up display and overall perceived situation awareness while using the technology. After the questionnaire, we instructed them to drive back to the starting location of the experiment, where they were dismissed. Throughout the study session, participants engaged in four experimental laps, with the session lasting between 1.5 to 2 hours.

## 4. MEASURES AND ANALYSIS

Where relevant, depending on the normalcy of the data, we performed analysis of variance to examine possible main effects and interactions among performance and survey metrics. Since this work is a within-subject design, each participant was subjected to all conditions. Therefore, repeated measures were collected per participant and cannot be considered independently. In this way, a mixed-methods model with a random effect μ was used to analyze the data to account for individual-specific heterogeneity. The level of significance is established at $p < 0.05$. Where relevant, Bonferroni was used for post hoc analyses. If normalcy was not achieved and data transformation could not correct the normalcy assumption, we used non-parametric analyses.

### 4.1 AR TASK PERFORMANCE

Using custom software, we computed the average duration time to complete AR tasks for each event. We also calculated the number of AR tasks participants completed within one event.

We define an *event* as the time window beginning with the initiation of data collection for detection tasks and AR HUD text message tasks, and ending based on specific visual cues. For central detection tasks, the event concludes when the third brake light of the lead vehicle turns off. For peripheral detection tasks, it ends when the front bumper of the participant's vehicle passes either the detection target or the child mannequin positioned along the roadside.

The starting point of each event on the smart road was determined through a series of dry runs. These sessions helped identify optimal locations where peripheral targets would not be visible to drivers from a long distance, while also allowing for relatively consistent event durations across trials. This calibration ensured both ecological validity and methodological consistency. An illustration of an event is provided in Figure 0-5.

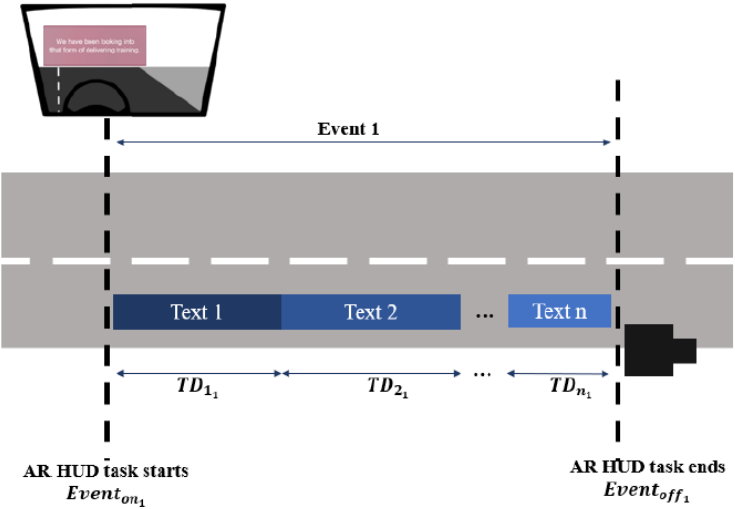

*Figure 0-5. An example of an event where the AR HUD task complexity is medium (two lines), and a participant is required to detect Adrian's small target on the side of the road. In this example, the participant was able to read three text messages with different durations (TDn).*

## 4.2 DETECTION DISTANCE

In this study, when a driver identified a specific target, the in-vehicle researcher activated a button within the vehicle. This action logged the driver's verbal response in the Data Acquisition System (DAS) and recorded the 'detection' variable. During the data processing phase, a pair of researchers employed 'Hawkeye,' a video data reduction tool developed by the Virginia Tech Transportation Institute. This software enabled researchers to simultaneously view multiple camera angles, facilitating precise determination of the time frame when the vehicle's front bumper passed the peripheral targets. This process was essential for accurately recording the 'passed' variable.

The 'detection' variable was also recorded using Hawkeye. This was due to feedback from the in-vehicle researcher about the potential inaccuracy of button presses during the data collection phase, attributed to the significant workload experienced. Subsequently, the team extracted GPS coordinate data from Hawkeye, encompassing both 'detection' and 'passed' variables. This data was crucial for calculating the distance at which participants detected the targets, measured in feet.

## 4.3 DETECTION PERFORMANCE

We calculated *hit rate* to assess drivers' responsiveness to detection stimuli within a specific time frame following stimulus onset, adhering to the ISO 17488:2016 [68] established threshold for response times between 100 ms and the upper limit in seconds. The upper limit varied based on stimulus duration and location: 2 seconds for the central stimulus and the moment the vehicle's front bumper passed peripheral stimuli. We corroborated verbal responses with video recordings from the driver assistance system shown in Figure 0-3. The hit rate, signifying driver response to a stimulus, was coded as a binary variable with "**yes**" (1) for a successful detection response and "**no**" (0) for non-responses or late detections beyond the upper threshold. For data analysis, we employed a generalized linear model (GLM) suitable for discrete data, utilizing a link function to associate the linear model with the non-normally distributed response. We implemented a binary logistic regression as the link function to model stimulus detection probability based on our exploratory variables.

## 4.4 INATTENTIONAL BLINDNESS

We assessed inattentional blindness by analyzing detection task performance and glance behavior metrics. We first coded participants' responses using two criteria: "*Did the driver respond to the stimulus on time?*" (no, delayed), and "*Did the driver look at the stimulus*" (yes, no). These four possible responses were used to create a "type of detection response" dependent variable, as shown in Table 0:2. In our analysis, eye movements landing within the Area of Interest (AOI) of a detection stimulus (i.e., the size of the lead vehicle brake light plus error margin) for a minimum of 100 ms were defined as stimulus fixations and labeled as 'looked'. Drivers' actions were coded as having responded to the stimulus when they verbally responded to it.

*Table 0:2. Classification of Detection Responses vs Eye Movements. Adapted from* [93]

| | | Detection Response | |
|---|---|---|---|
| | | NO | DELAYED |
| Eye movement (fixations) >100 ms | NO | **Did not look** at the stimulus<br>**Did not respond** to the stimulus<br>Ordinary Blindness, failed to look | **Did not look** at the stimulus<br>**Delayed Response** to the stimulus<br>Attention capture; Detected at the useful field of view |
| | YES | **Looked** at the stimulus<br>**Did not respond** to the stimulus<br>Inattentional Blindness | **Looked** at the stimulus<br>**Delayed Response** to the stimulus<br>Attention Capture; Detected at the fixation point |

## 5. RESULTS

### 5.1 AR TASK PERFORMANCE

We analyzed the AR Task Performance of 24 participants, each undergoing 18 AR task events, resulting in a comprehensive dataset of 432 events. Due to software issues, we lost 6 data points, leaving us with a final sample size of 426 data points for AR Task Performance analysis.

#### 5.1.1 NUMBER OF AR TASKS COMPLETED

In analyzing the average number of AR tasks participants were able to read aloud during a single event, we found the following results: On average, participants successfully completed 3.82 AR text messages of low visual demand per event (SD = 1.37; 95% CI [3.54, 4.10]), 2.26 AR text messages of medium visual demand per event (SD = 0.77; 95% CI [2.10, 2.41]), and 1.85 AR text messages of high visual demand per event (SD = 0.87; 95% CI [1.67, 2.03]). These findings, illustrated in Figure 0-6 with a 95% Bonferroni confidence interval plot, clearly indicate that as the visual demand of AR tasks increases, the average number of AR text messages participants are able to complete within the same event decreases significantly, as expected.

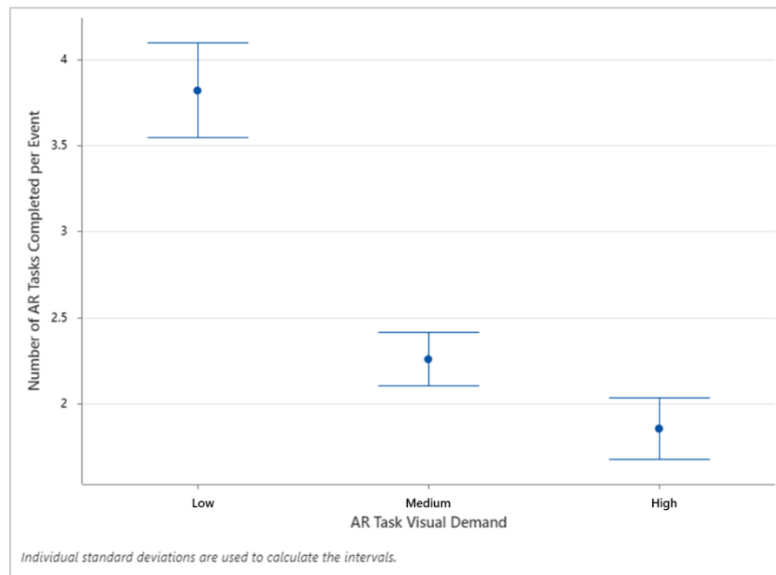

*Figure 0-6. 95% Bonferroni Confidence for the Mean Number of AR Tasks Completed Whitin One Event vs AR Task Visual Demand*

When analyzing the number of AR tasks completed within one event by participants based on the detection task stimulus, participants successfully completed, on average, 2.57 AR text messages (SD = 1.34; 95% CI [2.30, 2.84]) when detecting the brake 'light', 2.66 AR text messages (SD = 1.41; 95% CI [2.38, 2.94]) when detecting the 'child', and 2.71 AR text messages (SD = 1.28; 95% CI [2.45, 2.97]) when detecting the 'target'. These results, illustrated in Figure 0-7 with a 95%

Bonferroni confidence interval plot, suggest that participants' performance in completing AR tasks is slightly better when detecting the 'target' stimulus compared to the 'child' and 'brake light' stimuli. However, these differences do not appear to be statistically significant.

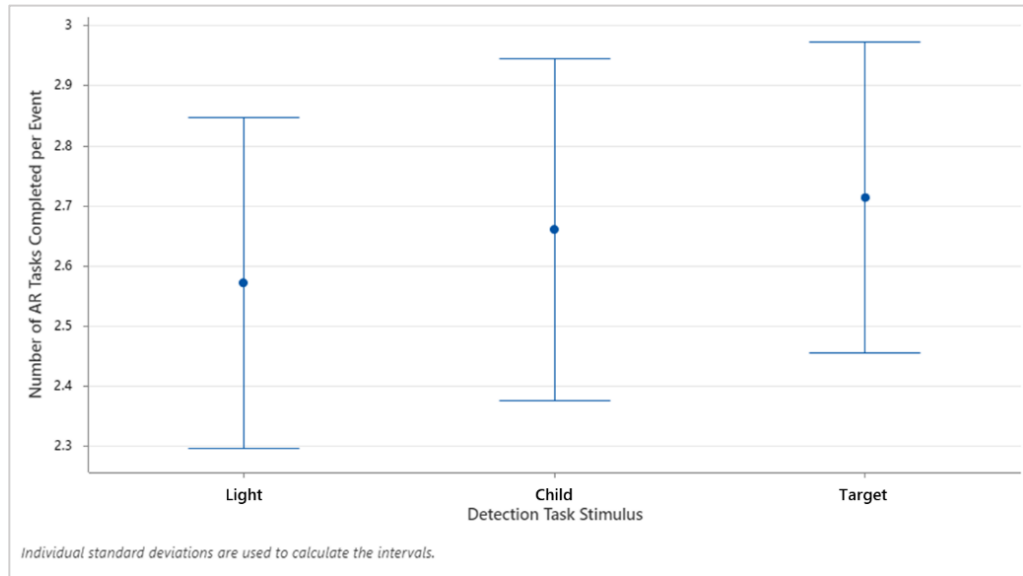

*Figure 0-7. 95% Bonferroni Confidence for the mean of the Number of AR Tasks Completed Within One Event vs Detection Task Stimuli*

We also analyzed the number of AR tasks completed by participants using a Poisson regression analysis with a natural log link function. The model explained 42.52% of the variance in task completion counts (adjusted $R^2$ = 39.53%), with an AIC of 1354.91, AICc of 1355.35, and BIC of 1391.43. Goodness-of-fit tests indicated a strong model fit (Deviance test: $\chi^2(418)$ = 154.10, p = 1.000; Pearson test: $\chi^2(418)$ = 164.36, p = 1.000). The regression equation identified significant negative coefficients for AR task of medium visual demand (-0.565, p < 0.000) and for AR task of high visual demand (-0.837, p < 0.000) compared to AR tasks of low visual demand, while detection task stimulus type and stimulus-visual demand interactions were non-significant. Completed results can be seen in

Table 0:3. The Wald test confirmed the overall model significance ($\chi^2(8)$ = 113.87, p < 0.000), with the AR Task Visual Demand variable being a significant predictor ($\chi^2(2)$ = 45.52, p < 0.000), but not the Detection Task Stimulus variable ($\chi^2(2)$ = 0.01, p = 0.993) or their interaction ($\chi^2(4)$ = 1.16, p = 0.885).

*Table 0:3. Poisson Regression Coefficient Table for Number of AR Tasks Completed Within One Event. Bold fonts indicate significant effects (p<0.05)*

| Term | Coef | SE Coef | Z-Value | P-Value | VIF |
|---|---|---|---|---|---|
| Constant | 1.3437 | 0.0737 | 18.23 | 0.000 | |
| Detection Task Stimulus | | | | | |
|   Child | -0.011 | 0.105 | -0.10 | 0.917 | 2.77 |
|   Target | 0.000 | 0.104 | 0.00 | 1.000 | 2.77 |
| AR Task Visual Demand | | | | | |
|   Medium | -0.565 | 0.125 | -4.52 | **0.000** | 3.58 |
|   High | -0.837 | 0.135 | -6.20 | **0.000** | 3.68 |
| Detection Task Stimulus* AR Task Visual Demand | | | | | |
|   Child * Medium | 0.035 | 0.175 | 0.20 | 0.841 | 2.95 |
|   Child * High | 0.176 | 0.185 | 0.95 | 0.341 | 2.96 |
|   Target * Medium | 0.078 | 0.173 | 0.45 | 0.653 | 3.04 |
|   Target * High | 0.154 | 0.186 | 0.83 | 0.408 | 2.90 |

### 5.1.2 DURATION OF AR TASK

To analyze how visual demand affects the time participants took to read a single AR text message, we observed the following average durations: 2.37 seconds for messages with low visual demand (SD = 0.65; 95% CI [2.27, 2.48]), 4.75 seconds for medium visual demand (SD = 0.77; 95% CI [4.60, 4.91]), and 7.28 seconds for high visual demand (SD = 0.87; 95% CI [6.28, 7.58]). These findings, illustrated in Figure 0-8 with a 95% Bonferroni confidence interval plot, clearly indicate that as the visual demand of AR tasks increases, the average time (in seconds) it takes for participants to read one single AR text message increases, as expected.

We also examined how reading duration varied depending on the detection task stimulus. On average, participants took 4.51 seconds (SD = 2.23; 95% CI [4.14, 4.89]) when detecting the 'brake light', 5.07 seconds (SD = 2.58; 95% CI [4.64, 5.50]) for the 'child', and 4.79 seconds (SD = 2.26; 95% CI [4.42, 5.15]) when detecting the 'target'. These results, illustrated in Figure 0-9 with a 95% Bonferroni confidence interval plot, suggest that participants' average time (in seconds) to read one AR text message is slightly higher when detecting the 'child' stimulus compared to the 'target' and 'brake light' stimuli.

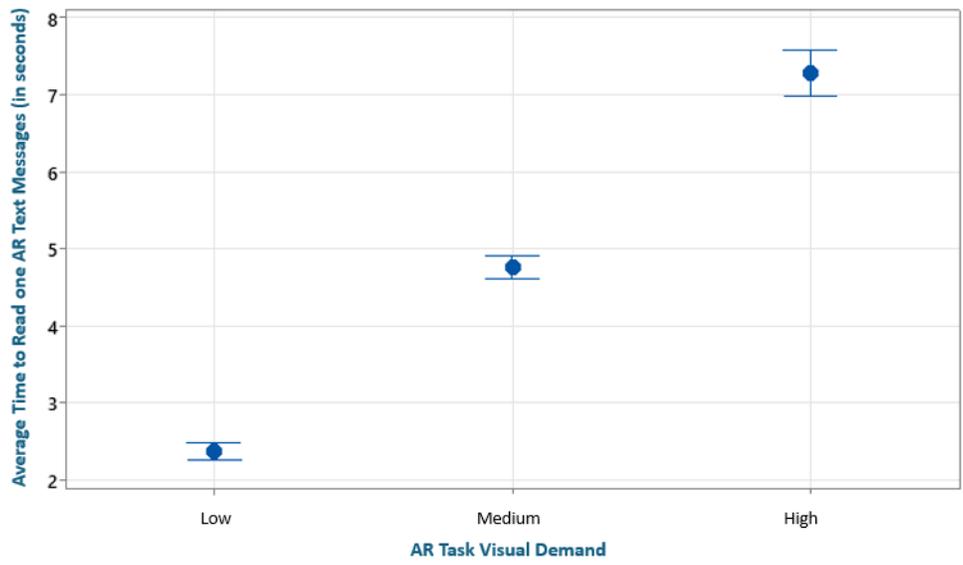

*Figure 0-8. 95% Bonferroni Confidence for Average Duration to Read an AR Text Message vs AR Task Visual Demand*

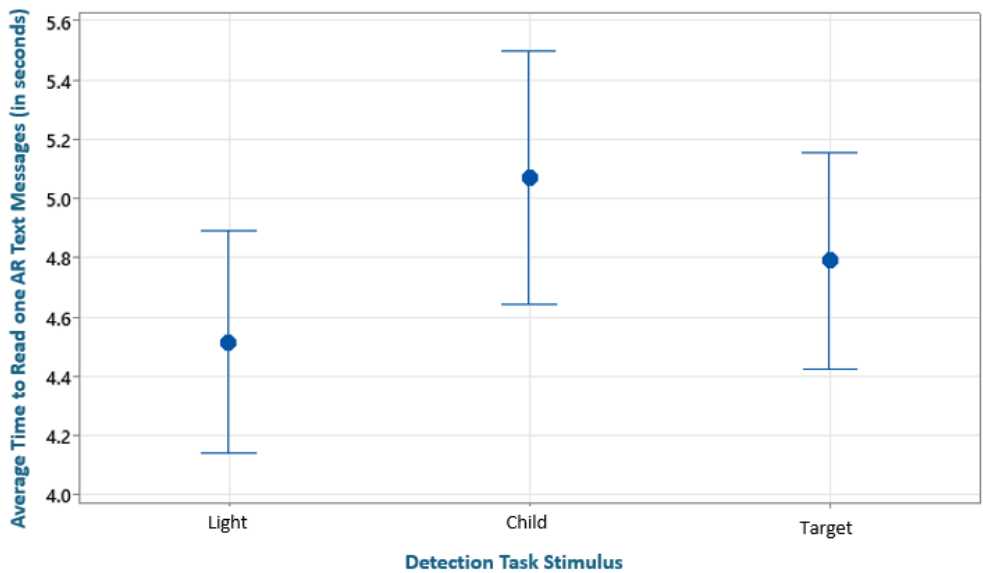

*Figure 0-9. 95% Bonferroni Confidence for Average Duration to Read an AR Text Message vs Detection Task Stimuli*

For the duration time (in seconds) it took to read a single AR text message, the mixed effects ANOVA revealed significant variability between participants (Z = 2.08, p = 0.019), accounting for 8.33% of the total variance, with the remaining 91.67% attributable to random error (Z = 14.03, p < 0.000). The model accounted for 77.05% of the variance in the average time to read AR Task texts (R-squared = 77.05%, adjusted R-squared = 76.61%). The fixed effects analysis showed (as shown in **Error! Not a valid bookmark self-reference.** that both Detection Task Stimulus F(2, 394.10) = 7.51, p = 0.001, $n_p^2 = 0.370$), and AR Task Visual Demand (F(2, 394.09) = 638.80, p < 0.000, $n_p^2 = 0.76$) significantly influenced the average time to read AR Task texts. However, the interaction between Detection Task Stimulus and AR Task Visual Demand was not significant (F(4, 394.10) = 1.07, p = 0.373, $n_p^2 = 0.005$). See Figure 0-10 for main effect plot of fitted means.

Table 0:4. Fixed Effects Table for Mixed Effects ANOVA of Average Duration to Read an AR Text Message. Bold fonts indicate significant effects (p<0.05)

| Term | DF Num | DF Den | F-Value | P-Value | $n_p^2$ |
|---|---|---|---|---|---|
| Detection Task Stimulus | 2.00 | 394.10 | 7.51 | **0.001** | 0.37 |
| AR Task Visual Demand | 2.00 | 394.09 | 638.80 | **0.000** | 0.76 |
| Detection Task Stimulus * AR Task Visual Demand | 4.00 | 394.10 | 1.07 | 0.373 | 0.005 |

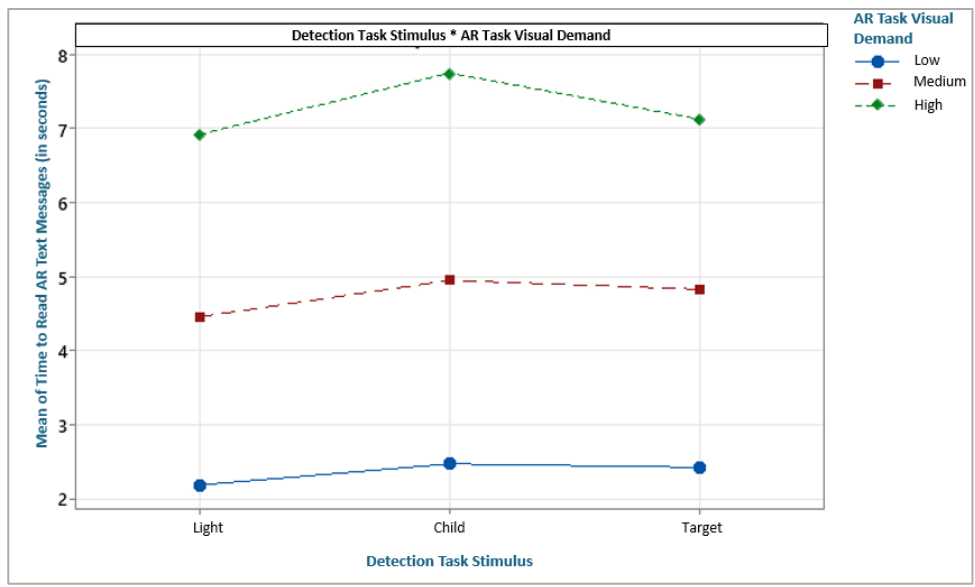

Figure 0-10. Main Effect Plot for Average Duration to Read an AR Text Message – Fitted Means.

The Bonferroni simultaneous tests for differences of means revealed significant differences in the average time it takes to read AR text messages across various levels of AR visual demand. Specifically, participants took an average of 2.383 seconds less to read AR text messages of low visual demand compared to those of medium visual demand ($T(394.075) = 17.31$, $p < 0.000$). Similarly, there was a significant decrease of 4.902 seconds to read AR text messages of low visual demand compared to those of high visual demand ($T(393.966) = 35.74$, $p < 0.000$). Additionally, the comparison between high and medium levels of AR visual demand showed a significant difference of 2.519 seconds ($T(394.236) = 18.23$, $p < 0.000$). These results indicate that as the visual demand of AR tasks increases, the time required to read the AR text messages also increases significantly, as it is illustrated on Figure 0-11.

Further pairwise comparisons showed a statistically significant increase in the average time spent reading AR texts when the detection task stimulus was a 'child' compared to the brake 'light' ($T(394) = 3.87$, $p < 0.000$). No statistical difference was found in the time spent to read AR texts when the detection task stimulus was a 'target' compared to the brake 'light' ($T(394) = 1.97$, $p = 0.149$) or the 'child' ($T(394) = -1.92$, $p = 0.168$). These findings suggest that the presence of a 'child' stimulus significantly increases reading time compared to a brake 'light', while the 'target' stimulus does not significantly differ from either, as it is illustrated on Figure 0-11.

*Table 0:5. Bonferroni Simultaneous Tests for Differences of Means – Average Duration to Read an AR Text Message. Bold fonts indicate significant effects (p<0.05)*

| Difference of Levels | Difference of Means | SE of Difference | DF | Simultaneous 95% CI | T-Value | Adjusted P-Value |
|---|---|---|---|---|---|---|
| Child – Light | 0.534 | 0.138 | 394.159 | (0.210, 0.859) | 3.87 | **0.000*** |
| Target – Light | 0.272 | 0.138 | 394.159 | (-0.053, 0.596) | 1.97 | 0.149 |
| Target – Child | -0.263 | 0.137 | 393.978 | (-0.585, 0.060) | -1.92 | 0.168 |
| Medium - Low | 2.383 | 0.138 | 394.075 | (2.059, 2.707) | 17.31 | **0.000*** |
| High - Low | 4.902 | 0.137 | 393.966 | (4.579, 5.224) | 35.74 | **0.000*** |
| High – Medium | 2.519 | 0.138 | 394.236 | (2.194, 2.844) | 18.23 | **0.000*** |

*Individual confidence level = 98.09%*

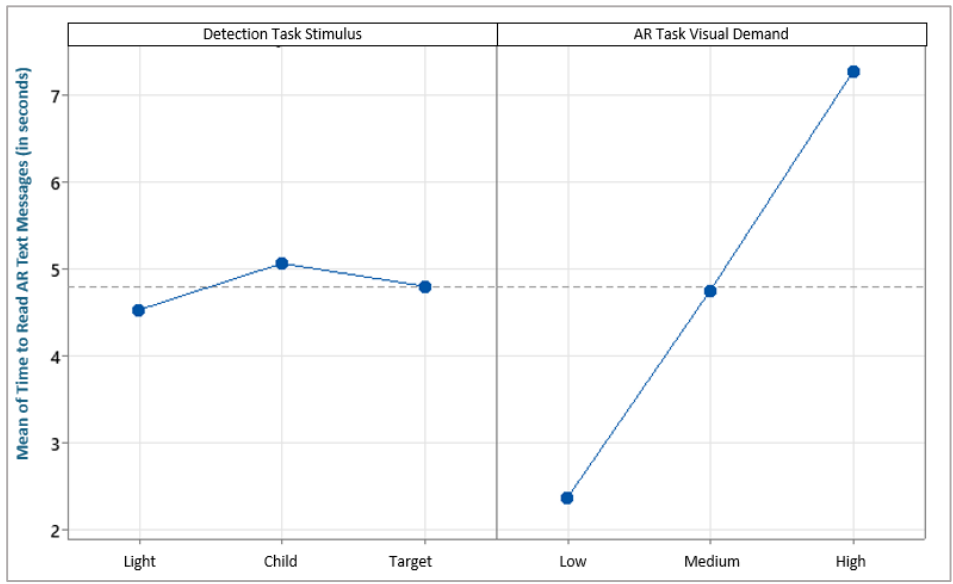

*Figure 0-11. Main Effect Plot for Average Time to Read an AR Text Message -Fitted Means.*

## 5.2 DETECTION PERFORMANCE

We analyzed the detection performance of 24 participants, each of whom experienced a total of 24 detection events, resulting in a comprehensive analysis of 576 detection events that were coded as binary 1,0 responses, as described in section 4.3. Our findings revealed that **87.5% of participants** (21 out of 24) either failed to respond or had a delayed response to at least one detection event. Additionally, **10.42% (60 out of 576)** of the detection events were either missed or experienced a delayed response. A more detailed analysis of these events is provided in section 5.4.

Using a generalized linear model with binary logistic regression as the link function, the Wald test confirmed the overall model significance ($\chi^2(5) = 29.41$, $p < 0.000$), with both the AR Task Visual Demand variable ($\chi^2(2) = 18.61$, $p < 0.000$), and the Detection Task Stimulus variable ($\chi^2(3) = 11.56$, $p = 0.009$) being significant predictors of the probability of detecting a stimulus on the environment. The main effect plot can be seen in Figure 0-12.

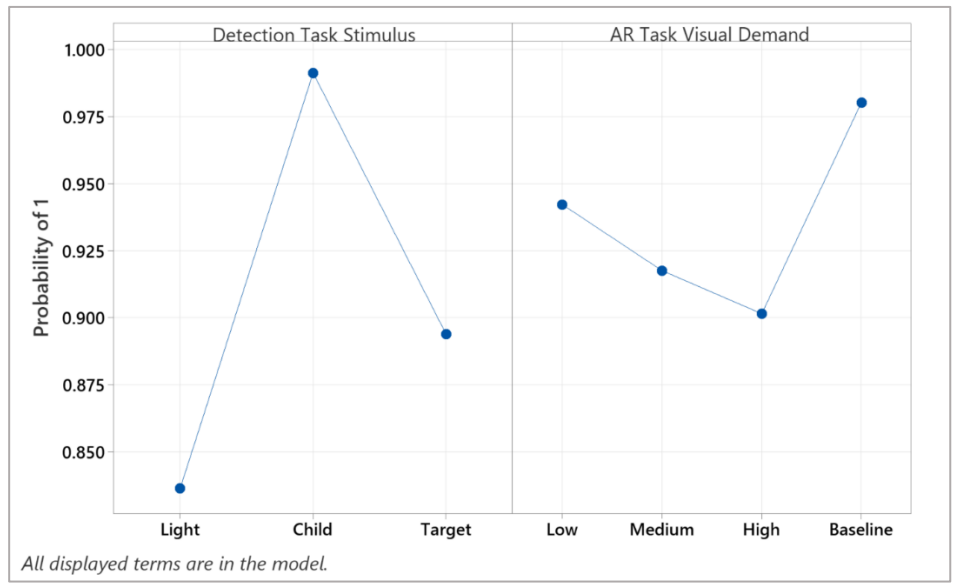

*Figure 0-12. Main effect plot showing the probability of event detection based visual demand of the AR Task and the stimulus of the detection task. As demonstrated, the presence of AR consistently decreases the likelihood of event detection, with a greater reduction observed as the visual demand of the AR task increases.*

The logistic regression model explained 14.18% of the variability in the outcome variable (Deviance R-squared = 14.18%), with an adjusted explained variability of 12.88% (Adjusted Deviance R-squared = 12.88%). The model's Akaike Information Criterion (AIC) was 342.35, while the corrected AIC (AICc) was 342.50, indicating a reasonable fit considering the model's complexity. Additionally, the Bayesian Information Criterion (BIC) value was 368.49, providing a more conservative estimate of the model's fit. The model demonstrated good discriminative ability, with an Area Under the Receiver Operating Characteristic Curve (AUC-ROC) of 0.7701, suggesting that it is effective in distinguishing between the positive and negative outcomes.

Participants who were not performing a secondary AR task (baseline condition) had statistically significantly higher odds of detecting a stimulus in the environment compared to those performing AR tasks with low (OR = 3.0443, 95% CI [1.0515, 8.8140]), medium (OR = 4.4549, 95% CI [1.5902, 12.4801]), or high (OR = 5.4161, 95% CI [1.9573, 14.9874]) visual demand. These results indicate that the presence of AR tasks decreases the probability of detecting stimuli in the environment compared to the baseline. Complete odds ratio results can be seen in Table 0:6.

Additionally, we found significant differences in the odds of detecting a stimulus in the environment when it was located in the periphery compared to the central field of view of drivers when detecting the 'child' (OR = 22.0365, 95% CI [5.1976, 93.4301]), but not for the 'target' (OR = 1.6476, 95% CI [0.9237, 2.9388]). These results suggest that people might attend more to objects for detection rather than spatial locations, as highlighted in [21]. Lastly, when examining stimuli located in the periphery, we found significant differences in the odds of detecting the 'target' compared to the 'child' mannequin (OR = 0.0748, 95% CI [0.0173, 0.3230]).

*Table 0:6. Odds Ratio for Detection Task Stimuli & AR Task Visual Demand variables. Bold fonts indicate significant effects (p<0.05)*

| LEVEL A | LEVEL B | ODDS RATIO | 95% CI |
|---|---|---|---|
| Detection Task Stimuli | | | |
| Child | Light | 22.0365 | **(5.1976, 93.4301)** |
| Target | Light | 1.6476 | (0.9237, 2.9388) |
| Target | Child | 0.0748 | **(0.0173, 0.3230)** |
| AR Task Visual Demand | | | |
| Medium | Low | 0.6834 | (0.3211, 1.4542) |
| High | Low | 0.5621; | (0.2686, 1.1760) |
| High | Medium | 0.8225 | (0.4127, 1.6391) |
| Baseline | Low | 3.0443 | **(1.0515, 8.8140)** |
| Baseline | Medium | 4.4549 | **(1.5902, 12.4801)** |
| Baseline | High | 5.4161 | **(1.9573, 14.9874)** |

## 5.3 DETECTION DISTANCE

During the data processing stage, several instances of missing data were identified. This was primarily due to researchers being unable to properly hear participants' voices because of static interference when they reported detecting a target. Additionally, sun glare in the video recordings made it impossible to verify when participants passed specific stimuli, thus preventing the calculation of detection distances. In addition, there were instances where participants could detect a target before a text was shown, and as such, these data points were also excluded. Consequently, the sample size for the 'detection distance' variable was significantly smaller than for other variables investigated in the study.

For the 'child' stimulus, our final data sample for detection distance consisted of 127 data points from 21 participants, representing 66.15% of the original 192 data points. For the 'target' stimulus, we analyzed a total of 119 data points, which is 61.98% of the original 192 data points. The 'light' stimulus posed the most challenges due to sun glare and the low quality of the forward video, which made it difficult for researchers to accurately determine when the brake light activated/deactivated to calculate the detection distance from the lead car. Consequently, the final dataset for the 'light' stimulus includes only 51 events. However, considering that 60 events of missed or delayed occurrences were excluded from the detection distance analysis, our final dataset consisted of 297 events out of the potential 516. This represents 57.76% of the total events that should have been included in the detection analysis.

In our analysis of the detection distance of stimuli under different levels of AR task visual demands, we observed the following results: Participants detected stimuli at an average distance of 124.73 feet (SD = 78.08; 95% CI [107.13, 142.34]) when engaged with AR text messages of low visual demand. For AR text messages of medium visual demand, the detection distance averaged 114.80 feet (SD = 84.80; 95% CI [94.13, 135.50]). When AR text messages were of high visual demand, the average detection distance was 114.03 feet (SD = 70.61; 95% CI [97.20, 130.87]). In the baseline condition, where no AR text was being read, participants detected stimuli at an average distance of 150.63 feet (SD = 89.97; 95% CI [130.86, 170.40]). Figure 0-13 illustrates the 95% Bonferroni confidence interval plot for detection distance relative to the visual demand of the AR tasks.

These results suggest that the presence of AR text messages, irrespective of their visual demand level, reduces the detection distance of stimuli compared to the baseline condition. This reduction is more pronounced with medium and high visual demand AR texts. The baseline condition, where no AR text was read, had the highest average detection distance, indicating that participants' ability to detect stimuli improved significantly in the absence of visual distractions from AR texts.

In our analysis of detection distance based on the type of stimulus, participants responded at varying distances: The average detection distance was 97.17 feet (SD = 53.90; 95% CI [78.48, 115.87]) for detecting the brake 'light'. When detecting the presence of a 'child', participants responded at an average distance of 154.22 feet (SD = 94.22; 95% CI [133.99, 174.51]). For detecting the 'target', the average detection distance was 111.03 feet (SD = 70.00; 95% CI [95.48, 126.62]). Figure 0-14Figure 0-14 presents the 95% Bonferroni confidence interval plot for detection distance relative to the type of stimulus being detected.

The results indicate that the detection distance varies significantly depending on the type of stimulus. Participants detected the child at the greatest distance, suggesting a higher sensitivity or priority for this type of stimulus. In contrast, the brake light was detected at the shortest distance, which might indicate lower sensitivity. The target's detection distance falls between these two extremes, showing a moderate level of responsiveness.

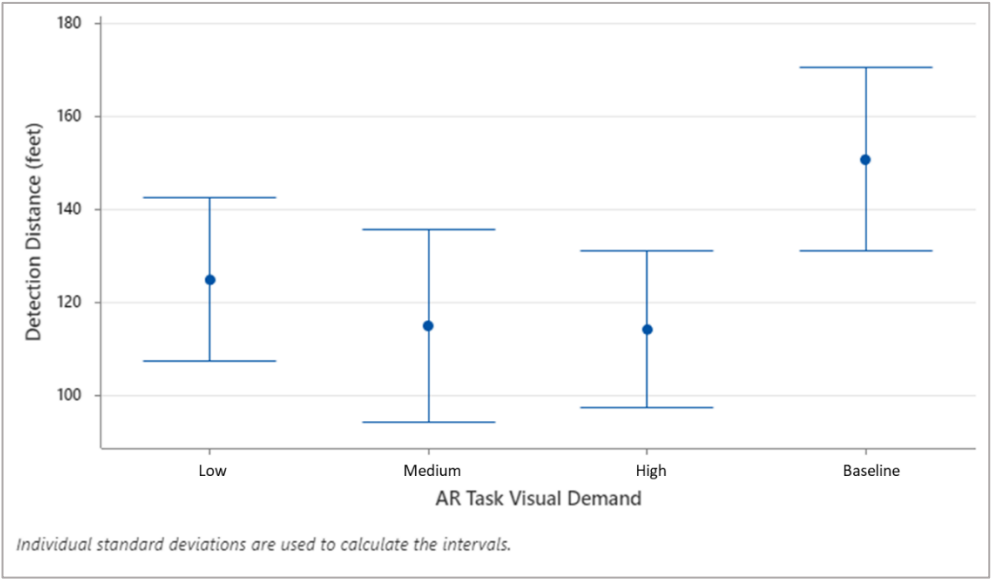

*Figure 0-13. 95% Bonferroni Confidence Interval for Detection Distance (in feet) vs AR Task Visual Demand*

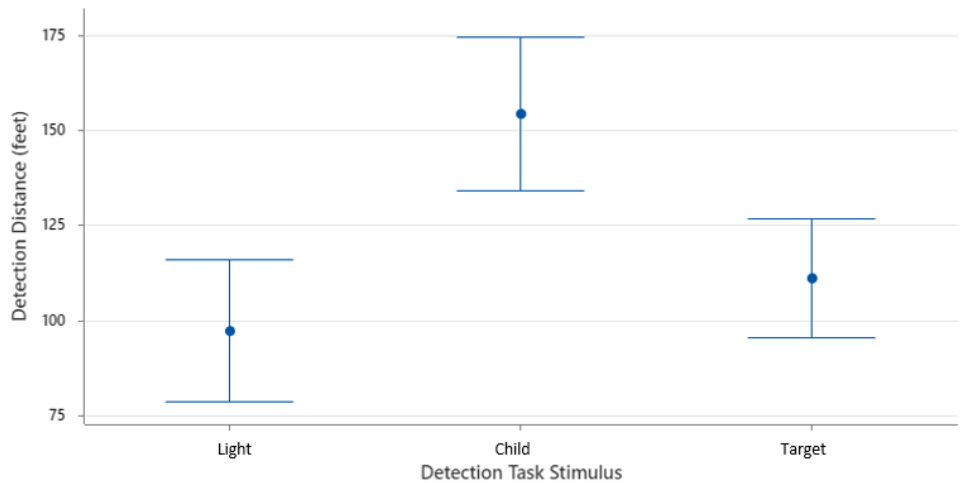

*Individual standard deviations are used to calculate the intervals.*

*Figure 0-14. 95% Bonferroni Confidence Interval for Detection Distance (in feet) vs Detection Task Stimulus*

The mixed-effects ANOVA results for detection distance revealed significant variability between participants (Z = 2.09, p = 0.018), accounting for 13.03% of the total variance, with the remaining 86.97% attributable to random error (Z = 11.63, p < 0.001). The fixed effects indicate that both AR Task Visual Demand (F(3, 273.77) = 4.63, p = 0.004, $n_p^2 = 0.048$), and Detection Task Stimulus (F(2, 286.29) = 16.71, p < 0.001, $n_p^2 = 0.105$) significantly influence detection distance. The model accounted for 26.68% of the variance (R-squared = 26.68%, adjusted R-squared = 25.42%).

For AR Task Visual Demand, post hoc analysis using Bonferroni Simultaneous Tests for Differences of Means revealed that the difference between medium and low was not significant (T(273.206)=-0.73, p = 0.886), nor was the difference between high and low (T(274.155)=0.652, p = 0.652). The comparison between the baseline and low indicated a non-significant increase in detection distance (T(272.667)=0.109, p = 0.109). The difference between high and medium was also non-significant (T(276.355)=-0.41, p = 0.977). However, the differences between baseline and medium (T(273.448)=2.91, p = 0.021) and between baseline and high (T(273.159)= 3.37, p = 0.005) were significant. For Detection Task Stimulus, the Bonferroni post hoc test showed significant differences between 'child' and 'light' (T(290.967)=4.82, p < 0.001) and between 'target' and 'child' (T(290.988)=1.42, p < 0.001), while the difference between 'target' and 'light' was not significant (T(277.653)=-4.67, p = 0.330). Complete post-hoc results can be seen on Table 0:7.

*Table 0:7. Bonferroni Simultaneous Tests for Differences of Means – Detection Distance. Bold fonts indicate significant effects (p<0.05)*

| Difference of Levels | Difference of Means | SE of Difference | DF | Simultaneous 95% CI | T-Value | Adjusted P-Value |
|---|---|---|---|---|---|---|
| Medium – Low | -8.9 | 12.2 | 273.206 | (-41.4, 23.6) | -0.73 | 0.886 |
| High – Low | -14.0 | 12.1 | 274.155 | (-46.2, 18.1) | -1.16 | 0.652 |
| Baseline - Low | 26.2 | 11.6 | 272.667 | (-4.6, 57.0) | 2.26 | 0.109 |
| High - Medium | -5.2 | 12.6 | 276.355 | (-38.7, 28.4) | -0.41 | 0.977 |
| Baseline – Medium | 35.1 | 12.1 | 273.448 | (3.0, 67.3) | 2.91 | **0.021** |
| Baseline - Hard | 40.3 | 11.9 | 273.159 | (8.5, 72.0) | 3.37 | **0.005** |
| Child - Light | 62.9 | 13.0 | 290.967 | (31.5, 94.3) | 4.82 | **0.000** |
| Target - Light | 18.8 | 13.2 | 290.988 | (-12.9, 50.5) | 1.42 | 0.330 |
| Target - Child | -44.13 | 9.44 | 277.653 | (-66.87, -21.40) | -4.67 | **0.000** |

*Individual confidence level = 98.09%*

## 5.4 INATTENTIONAL BLINDNESS

When examining inattentional blindness, we chose to distinguish between events that were completely missed and those that were detected after a delay. We made this distinction because delayed detection provides a better indicator of attention capture when using augmented reality, whereas inattentional blindness was used to assess events that were missed when drivers were looking directly at them. Out of the initial 60 missed or delayed detection events, 30% (18 events) were classified as delayed events, while 70% (42 events) were classified as fully missed events.

In the AR task with "low" visual demand, participants missed 9 events and exhibited delayed responses to 5 events. At the "medium" visual demand level, 14 events were missed, and 5 events experienced delayed responses. For the "high" cognitive load level, participants failed to identify 15 events and demonstrated delayed responses to 7 events. These results indicate that the number of fully missed and delayed detected events increased with the AR task visual demand, as previously shown by detection performance analysis. A visual breakdown of these results can be seen in Figure 0-15 and Figure 0-16.

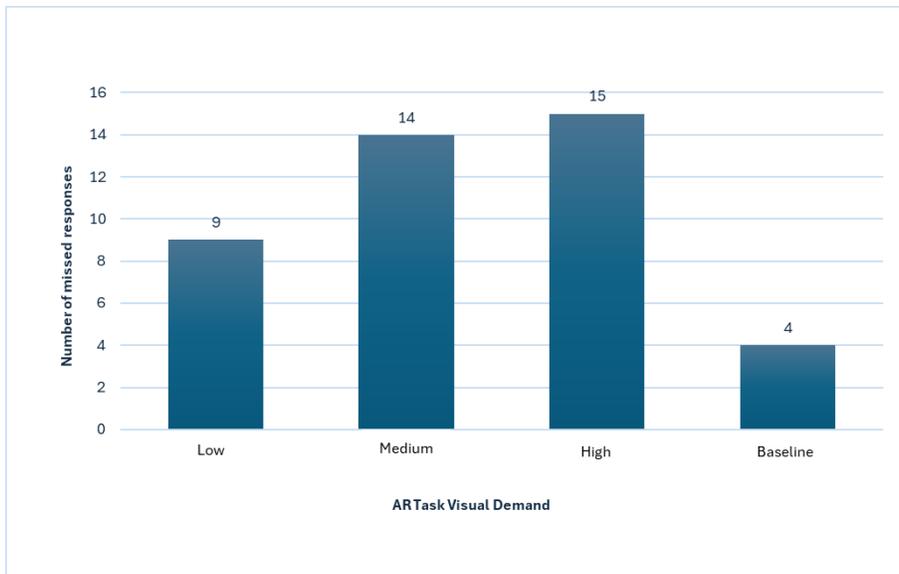

*Figure 0-15. Bar plot showing missed responses to detection events by AR task visual demand*

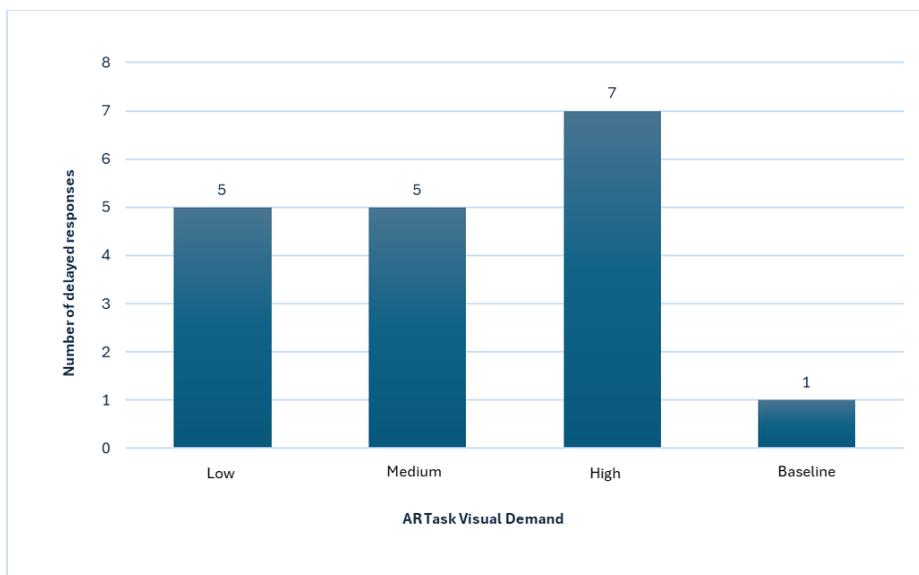

*Figure 0-16. Bar plot showing delayed responses to detection events by AR task visual demand.*

Additionally, when examining events within the drivers' central field of view (e.g., brake light), we found that the majority of these events were entirely missed (91.42%). In contrast, for events situated at the periphery of the drivers' field of view (e.g., target and child), most resulted in delayed detection responses (60%) rather than being fully missed. Notably, there were no instances where drivers completely failed to detect a child on the roadway.

In order to ascertain whether drivers failed to notice or reacted late to events on the road due to lack of visual attention we conducted a thorough analysis of the participants' gaze patterns during these detection events. Specifically, we examined each participant's fixation location for each missed or delayed event (as described in Table 0:8). Although eye tracking data was unavailable for five events due to gaze sampling issues, we were able to include a total of 55 events in our analysis using eye glance.

Interestingly, in 100% of the cases involving delayed events (17 events), drivers were found to be glancing both central and peripheral stimuli. This suggests that while attention capture occurred in these situations, drivers were still able to detect a stimulus at the fixation point.

As shown in Table 0:8, we characterize inattentional blindness as the complete failure to detect an event despite directly fixating on it. In our study, we identified 33 instances that qualified as inattentional blindness events. As demonstrated in Figure 0-17 and corroborated by detection performance metrics, the prevalence of inattentional blindness increases in tandem with the visual demand of the AR secondary task. Notably, this phenomenon is more pronounced within the drivers' central field of view (87.88%; 29 events) compared to drivers' peripheral field of view (12.12%; 4 events, 'target'). No occurrence of inattentional blindness was reported when participants needed to detect the 'child' in the side of the roadway.

*Table 0:8. Classification of the 55 Detection Responses vs Eye Movements Analyzed in this Study.*

|  |  | Detection Response | |
|---|---|---|---|
|  |  | NO | DELAYED |
| Eye movement (fixations) >100 ms | NO | **Did not look** at the stimulus<br>**Did not respond** to the stimulus<br>Ordinary Blindness, failed to look<br>(5 events) | **Did not look** at the stimulus<br>**Delayed Response** to the stimulus<br>Attention capture; Detected at the useful field of view<br>(0 events) |
| | YES | **Looked** at the stimulus<br>**Did not respond** to the stimulus<br>Inattentional Blindness<br>(33 events) | **Looked** at the stimulus<br>**Delayed Response** to the stimulus<br>Attention Capture; Detected at the fixation point<br>(17 events) |

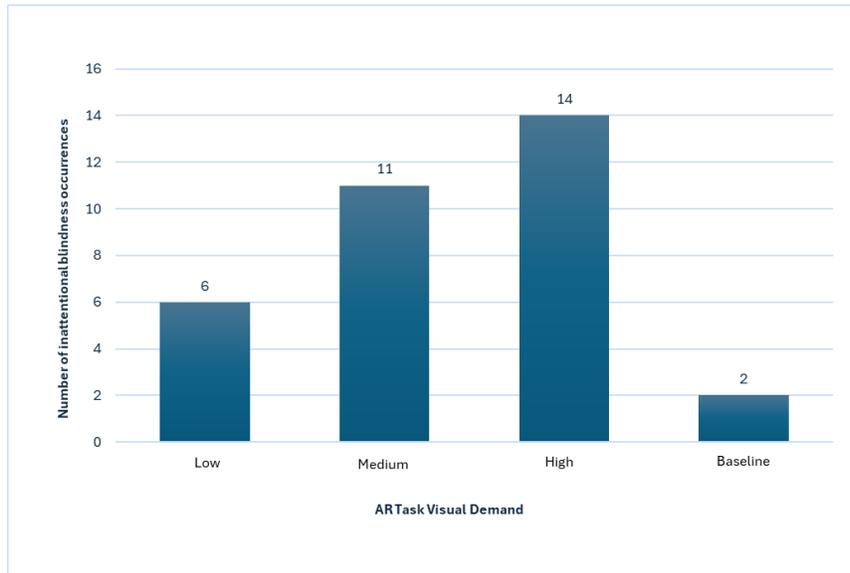

*Figure 0-17. Bar Plot of Number of Inattentional Blindness Occurrences by AR Task Visual Demand.*

## 5.5 SELF REPORTED PERCEPTIONS

We asked participants to complete a post-trial questionnaire after finishing the experimental session, in which they provided feedback on their perceptions of the head-up display and their overall situation awareness while using the technology. The results indicated that 80% of participants would not consider a head-up display safe to use while driving, and only 12% had a positive attitude towards its safety (see Figure 0-18 for a detailed breakdown of safety perceptions). Additionally, when asked if they would commonly use the HUD they experienced, 68% of participants responded with "not at all" or "rarely" (refer to Figure 0-19 for more details).

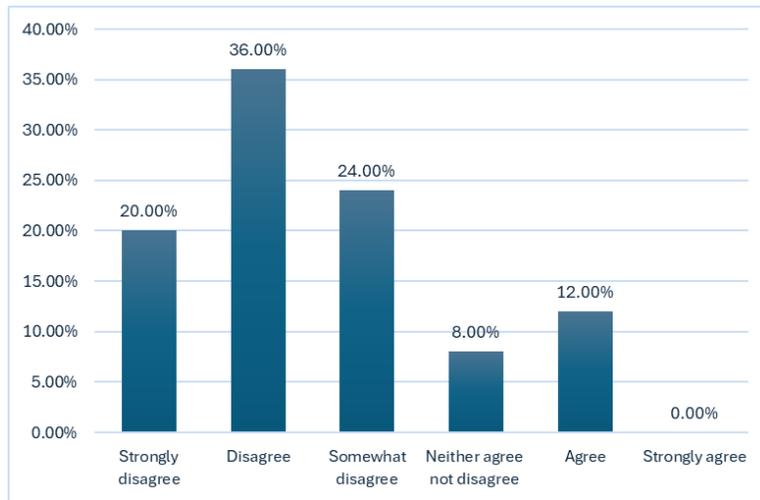

*Figure 0-18. Bar Plot of "Would You Consider The Head-Up Display (HUD) You Experienced Is "Safe To Use" Whilst Driving?"*

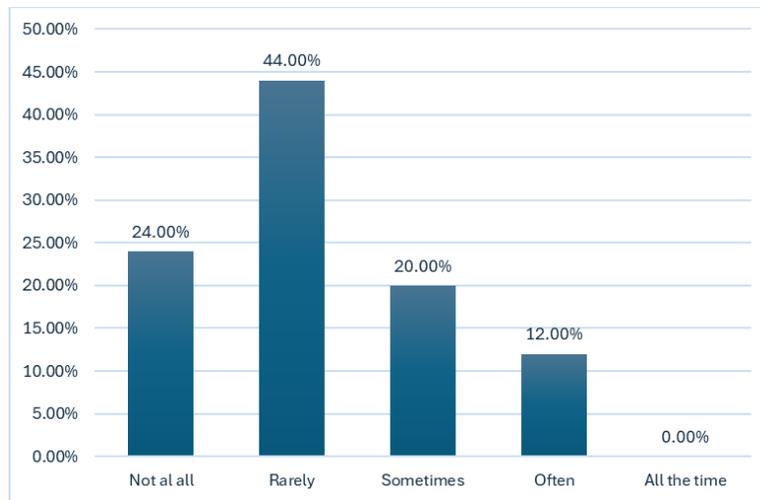

*Figure 0-19. Bar Plot Of "Would You Commonly Use The HUD That You Experienced?"*

Interestingly, when asked, "Did you feel aware of your surroundings?", a significant portion of the participants, 48%, reported feeling aware of their surroundings "Sometimes." This suggests that nearly half of the participants experience situational awareness intermittently. Following this, 36% of participants indicated they felt aware "Often," highlighting a considerable number of individuals who regularly maintain a heightened sense of their environment. In contrast, 8% of respondents each selected "Rarely" and "All the time," suggesting that a smaller fraction of the participants felt that they experienced low or constant awareness. Notably, no respondents selected "Not at all," implying that all participants felt some degree of awareness at least occasionally (see Figure 0-20).

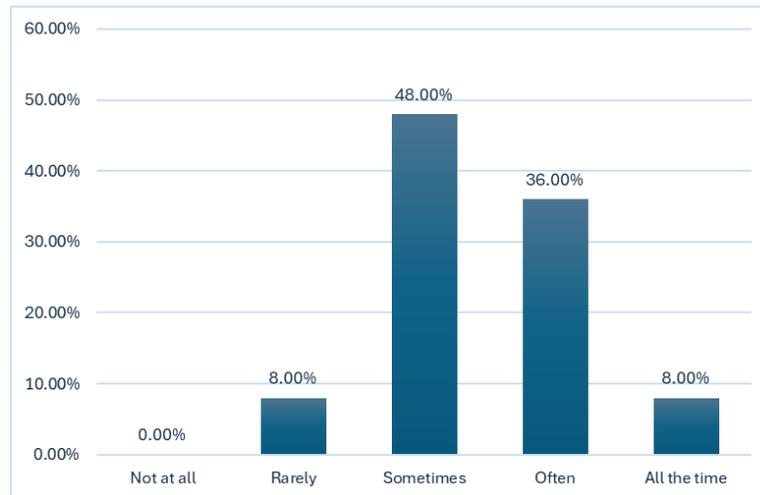

*Figure 0-20. Bar plot of "Did you feel aware of your surroundings"*

The majority of participants, 56%, reported feeling aware of the lead car "Often" when its rear lights were activated. This indicates that over half of the participants self-reported they felt that they frequently noticed the lead car's rear lights, suggesting a high level of situational awareness in these instances. Following this, 24% of respondents indicated they felt aware "Sometimes," showing that nearly a quarter of the participants experienced intermittent awareness of the lead car's rear lights. Additionally, 12% of participants reported feeling aware "All the time," which represents a smaller but significant portion of individuals who consistently noticed the rear lights. Conversely, 8% of respondents indicated they "Rarely" felt aware of the lead car's rear lights, and notably, none of the respondents selected "Not at all," implying that all participants had some degree of awareness when the rear lights were turned on (see Figure 0-21).

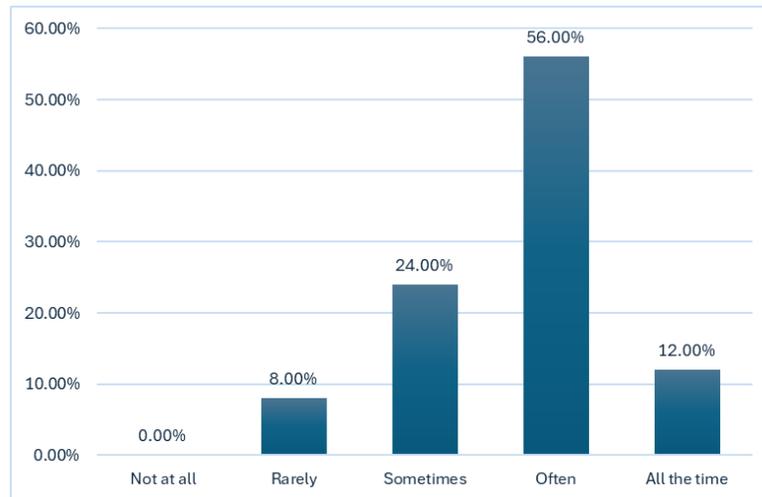

*Figure 0-21. Bar plot of "To what degree did you feel aware of the lead car when it turned its rear lights on?"*

A substantial majority of participants, 64%, reported feeling aware of stimuli on the sideroad "Often", indicating a high level of situational awareness in these scenarios. Additionally, 28% of participants indicated they felt aware "Sometimes," reflecting that a significant portion of the participants experienced occasional awareness of stimuli on the sideroad. Furthermore, 4% of respondents reported feeling aware "All the time," representing a smaller group of individuals who consistently self-reported they noticed sideroad objects. Conversely, 4% of respondents selected "Rarely," indicating a minimal level of awareness for some participants. Notably, no respondents chose "Not at all," implying that all participants experienced at least some degree of awareness of siderroad objects (see Figure 0-22)

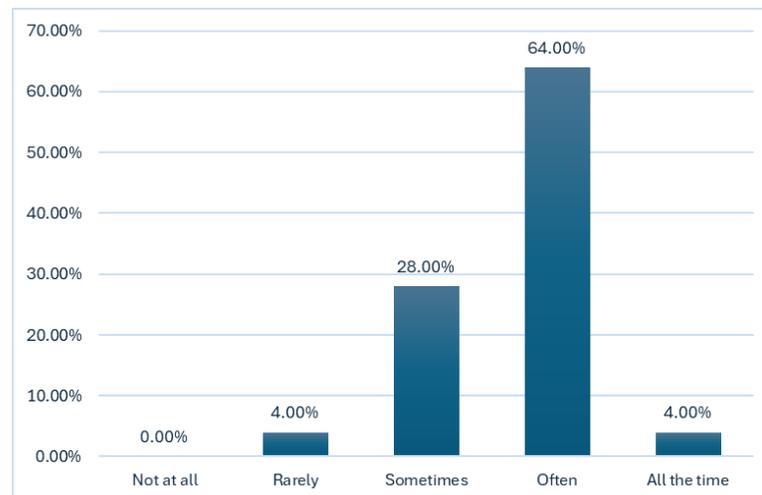

*Figure 0-22. Bar plot of "To what degree did you feel aware of objects on the sideroad??"*

## 6. DISCUSSION

The increasing use of in-vehicle AR technologies has raised concerns about driver safety. AR graphics displayed via HUDs (or other medium) can potentially distract drivers by challenging their visual focus and attention, making it crucial to consider the potential costs associated with reacting to critical hazards in the environment. In our study, we observed no unsafe driving behaviors, as drivers consistently managed to keep their vehicles within lane boundaries, indicating that primary task performance was not degraded to a level of unsafe behaviors with the use of AR. In fact, AR has been shown to afford better performance of vehicle control [94] and increased time with eyes on the road (due to the reduced scanning costs) [94] compared to traditional in-vehicle head-down displays. This is due to the fact that AR displays graphical information directly on the windshield, within drivers' forward field of view, allowing them to consume information while relying on ambient vision to maintain vehicle control. It is worth noting that while the benefits of AR in driving have been extensively studied, the assessment of their effectiveness in tertiary-task situations, such as the one presented in our study, remains limited.

We hypothesized that AR task visual demand influences the detection likelihood of stimulus on the roadway **(H1).** Our results support **H1** as drivers' ability to timely detect stimuli in the environment decreased as the AR task visual demand increased demonstrated by both detection performance **(H1A)** and inattentional blindness metrics **(H1B)**. A plausible explanation for these findings lies in the fact that drivers were simultaneously engaged in three distinct tasks: the primary driving task, an AR-based task, and a detection task. **The multiple resource theory** [95], [96] posits that multitasking is heavily constrained by the finite cognitive resources available for allocation to any given task. Consequently, tasks that are less cognitively demanding or require minimal effort can be more effectively shared with other concurrent tasks. As the visual demand of the AR task increased, drivers experienced a reduction in cognitive resources available for detecting stimuli in the outside environment. This, in turn, led to a higher incidence of inattentional blindness. Nonetheless, one may wonder why the driving task remained 'unaffected' by the increasing visual demand imposed by the AR task. To address this, it is essential to note that during the design phase of the AR task for this study, we conducted a user study utilizing a driving simulator. The primary aim was to ascertain that the visual demand at each level were distinguishable from one another, while simultaneously avoiding the elicitation of unsafe driving behaviors. In other words, our objective was to ensure that drivers could effectively multitask on the AR tasks without causing them to exceed beyond lane boundaries or result in collisions. Furthermore, drivers are often able to maintain their vehicle's position within the lane quite effectively due to the utilization of separate visual channels. By relying on ambient vision for lane centering, drivers can simultaneously glance at the AR display to read text messages, which necessitates focal vision [6]. This division of visual resources, known as **multiplicity of resources**, enables them to perform both tasks with relative efficiency. Overall, these findings emphasize the importance of understanding the resources being employed (*processing stages, processing code,* *perceptual modality, and visual subsystems,* read [20] for more details) during the design of AR experiences. As our cognitive resources are limited, multitasking is best supported when different resources are engaged [97]. For example, since driving primarily relies on visual-manual resources, it is crucial for AR interactions to avoid incorporating manual input to prevent significant interference with the driving task. In our study, we observed inattentional blindness even with a relatively simple AR user interface, which lacked elements such as animations or moving 3D objects that could potentially demand more attention. Additionally, our interface did not require any user interaction. We anticipate that inattentional blindness and attention capture could be exacerbated by more complex AR user interfaces. Further research is needed to investigate the effects of multimodal presentation of information and interaction modes in conjunction with AR on inattentional blindness.

Interestingly, inattentional blindness caused by AR displays appears to be more prevalent within drivers' central field of view. Our initial hypothesis **(H2A)** suggested that stimuli in the peripheral field of view would be more prone to being overlooked compared to those in the central field of view. This assumption was based on previous research in surface transportation, which demonstrated that drivers performing secondary tasks exhibited reduced sensitivity to peripheral events [35]–[38], leading to a higher frequency of "looked-but-failed-to-see" errors. A similar outcome, where inattentional blindness was more common in the periphery, has also been observed in the cognitive psychology domain [98]. Contrary to our expectations, our findings do not support these assumptions. We observed that even when participants' gaze direction was directly on the stimulus of interest (e.g., brake light within their central field of view), many failed to notice it. This finding corroborates with past research that shows that although there is a relationship between where people fixate and where they attend to [14]; the phenomenon of inattentional blindness illustrates that attention and fixation can be separated from each other [15].

It is important to highlight that the advantages of AR displays are postulated to be influenced by the **information integration** within near and far domains [20], [50]. This process encompasses mental information integration, wherein attention must be divided among multiple elements, but both contribute to a single task (cognitive or motor response). Consequently, the combined effects of these elements must be mentally integrated. In our study, reading a text message while driving and identifying potential hazards constitutes a dual task, divided attention scenario. In this context, display elements are associated with distinct responses and goals, separate from hazard detection and driving. Although AR graphics were superimposed onto the driving scene and the central detection stimulus, the absence of information integration between the AR task, driving task, and detection task may have contributed to the instances of inattentional blindness observed in our results. As such, our findings suggest an important design guideline for evaluating HUDs with safety in mind: <u>it is essential to ensure that HUD information is integrated with the driving task to potentially reduce occurrences of inattentional blindness.</u> Further research is needed to evaluate different types of everyday driving-related information and their degree of integration with the likelihood of inattentional blindness.

On the other hand, it is crucial to recognize that future AR applications in everyday life might not always be designed to fully integrate AR tasks with a primary task. For instance, individuals may use AR displays via smart glasses (or other medium) to check social media, send texts, and perform other activities while walking, driving, or working. Therefore, it is important to understand how the lack of integration between the primary task and the AR task affects the likelihood of inattentional blindness. We should consider interface design strategies (or even automation capabilities) that might mitigate the occurrence of this phenomenon. This aspect is especially relevant because displaying non-driving related information in front of drivers' line of sight, such as in this study, can lead drivers to believe they can attend to both the AR task displayed via HUDs and the real world simultaneously. When text messages are displayed on the center dashboard interfaces or cell phones, drivers are usually more conscious of the time spent looking away from the road and make an effort to refocus on their surroundings. In our study, when we asked participants, "*Did you feel aware of your surroundings?*", "*To what degree did you feel aware of the lead car when it turned its rear lights on?*", and "*To what degree did you feel aware of objects on the side road?*", only a small fraction (8%, 8%, and 4%, respectively) indicated they rarely felt aware of other elements when using the HUD. This indicates an over trust in their situational awareness performance with the technology.

Another highly relevant explanation for the numerous instances of inattentional blindness observed within drivers' central field of view may be linked to the **non-conformality** of our interface. With conformal imagery, the synchronized motion of near domain symbology and its far domain counterpart "connects" the two elements as the head or vehicle rotates, thus fostering parallel processing of both components as if they belong to a single object. This connection known as the **object-based theory** [45], [46], which may shed light on the prevalence of inattentional blindness in our study. Indeed, early research has demonstrated that humans cannot simultaneously process the world and a screen-relative display, perceiving them as two distinct images [45], [46], [99]. It is posited that a perfectly world-relative display might eliminate the time cost of switching between tasks and reduce the likelihood of inattentional blindness [99]. However, if graphics are poorly designed, rendered, and not precisely spatially registered with their real-world counterpart, users will perceive the interface as a screen-relative display. Achieving a truly world-relative commercial AR display (where every AR graphics is presented at accommodative depth that matches its real-world referent object) has not yet been accomplished, and AR images may still be processed separately from the world, implying that there might always be some cost associated with switching tasks.

Broadly speaking, our findings from **H2A** are consistent with prior research, suggesting that integrating AR into or overlaying it on tasks may increase the likelihood of people overlooking critical cues as they divide their attention between tasks occupying the same visual space in both driving [54], [60], and aviation [8], [100] domains. However, this work did not investigate whether the phenomenon would be different in the central and peripheral field of view of users. Also, it is important to note that in the aviation domain research, there was no change in visual accommodation as the display was collimated to optical infinity, and real-world referents are very far from the observer. In contrast, in our study, the AR task was presented at a different focal distance from the stimulus being detected. Previous research has shown that there are accommodation switching costs associated with the use of AR [49] which we believe may also have contributed to the occurrences of inattentional blindness observed in our study.

In examining **H2B**, we hypothesized that stimuli with lower perceived value would be more likely to be missed by drivers compared to those with higher perceived value. Our findings support **H2B**, as drivers failed to detect small targets more frequently than the child-sized mannequin. While the apparent salience of different stimuli could be a factor - as demonstrated by the child-mannequin being detected a further distance than the small target - our results can also be interpreted as reflecting the **resource allocation policy** [95] of attention during divided attention tasks. Drivers seem to prioritize lane-keeping and hazard monitoring of perceived real danger (such as the child mannequin) over detecting smaller targets with no significant intrinsic value. Consequently, in such situations, they allocate their finite cognitive resources to tasks they perceive as more critical. However, a key factor influencing this allocation policy is engagement [101]. Engaging and captivating tasks tend to receive the majority of resources when shared over time, often at the expense of concurrent tasks that may, in reality, be of higher importance—such as detecting a pedestrian crossing the street. In our study, some drivers might have been more engaged in reading text messages, prioritizing this activity over the more safety critical task of detecting stimuli in the roadway. Future studies should investigate stimuli of different perceived values at comparable levels of apparent salience to understand their effect on inattentional blindness. Additionally, further research is needed to predict and assess engagement levels with AR in situations where it may jeopardize safety. Developing methods to accurately measure and manage engagement could be crucial in ensuring that AR does not compromise users' ability to prioritize critical tasks and maintain safe behaviors.

Overall, while only 10.42% of events in our study were categorized as late or missed, the significance of our findings extends beyond mere statistical considerations. Conventional statistical techniques, which typically focus on analyzing means, may not adequately capture the safety implications of inattentional blindness and attention capture, as these phenomena can lead to unpredictable and potentially disastrous consequences. Accidents and unsafe conditions, which are the primary focus of our research, are not common events in certain contexts and should not be considered significant only if they occur frequently within a study. For example, it can be argued that a collision with any single pedestrian is undesired even if the data showed this collision was not statistically significant! Therefore, we believe our research highlights the need to address inattentional blindness when using AR and shows that we need new ways of thinking about the significance of results, beyond traditional statistics, to make sure emerging technologies are safe.

Finally, the crux of inattentional blindness research lies in truly unexpected events. These investigations delve into the propensity of information to capture attention when individuals have no prior expectation of the event's manifestation. Once participants became cognizant of the unexpected event, object, or sound, they may start anticipating the unanticipated, consequently altering the task's fundamental nature [102]. As such, inattentional blindness is thought to occur exclusively in truly unexpected situations [20]. In our study, drivers were informed that stimuli for detection would be located in either their central or peripheral field of view, and they were aware of the nature of these stimuli. Thus, these events cannot be classified as truly unexpected. Nonetheless, we observed numerous instances of both inattentional blindness and attention capture, suggesting that while expectancy influences the phenomenon, inattentional blindness also happens when individuals anticipate events in their surroundings. Regardless, we believe that our findings would be even more striking in scenarios where individuals held no expectations regarding events on the roadway.

## 6.1 IMPLICATIONS TOWARDS A SAFETY-CENTRIC EVALUATION FRAMEWORK FOR AR HUDS

Focal visual channels depend heavily on foveal vision to perform tasks that require discriminating fine details, such as reading text messages on AR displays, while ambient vision relies primarily on peripheral vision to perceive orientation and ego motion. In our study, drivers effectively utilized both focal and ambient vision to read AR text messages and maintain vehicle lane position. However, our results also demonstrated that when foveal vision is required for two tasks simultaneously, such as the detection task and AR reading task in our study, competition for focal resources arises, impacting the availability of these resources for identifying road hazards. Therefore, when evaluating AR systems in applications where safety is of utmost importance, it is crucial to consider the potential trade-offs and prioritize safety metrics. Traditional metrics, such as lateral driving performance decrement, may not be sensitive enough to assess the impact of HUDs on driver performance. When the workload is balanced between primary and secondary tasks, drivers might still effectively use their ambient vision to maintain lateral control of the vehicle. As such, the competition for cognitive resources between tasks requiring focal vision and occupying the same visual field of view, makes hazard detection methods more relevant and sensitive for evaluating the safety implications of HUDs on driver performance.

When assessing hazard detection during HUD evaluations, the advantages of augmented reality over existing technologies may not be as apparent as initially presumed, not only in the surface transportation domain but also in other domains where AR is used. This issue becomes more challenging when considering the standard methods, the AR community employs to evaluate AR systems and interfaces. A recent systematic review spanning a decade of augmented reality usability studies [103] revealed that the most prevalent metrics for AR usability studies are subjective ratings of preference (57%), followed by error/accuracy measures (45%) and task completion time (42%). Furthermore, the majority of user studies took place indoors (83%), as opposed to outdoors (15%) or a combination of both (2%). This is particularly concerning given that many of these studies focus on areas such as medicine (15%), perception (18%), industry (10%), navigation and driving (9%), and interaction (23%), where safety is a major concern. In our research, we examined a screen-relative text-message AR interface in a dynamic driving outdoor setting and demonstrated that drivers fail to detect some hazards even when directly looking at them. Our findings underscore crucial questions regarding the sufficiency of traditional AR evaluation metrics, which might not reveal safety concerns. Furthermore, it is essential to recognize that future AR applications are more likely to support primary tasks such as assembly, surgery, or driving, rather than being the primary task itself. Consequently, while performance and user experience with AR are important considerations at different stages of the product development lifecycle, they should not be the sole focus and metrics for certain applications where safety is critical.

As a reminder, our proposed detection approach aimed to improve ecological validity compared to traditional signal detection tasks, as our study was conducted on a real roadway. We created more salient stimuli relevant to driving that did not interfere with the driving task, avoiding redirection of the driver's attention. Additionally, the regular and less frequent presentation of our stimuli made them less predictable for drivers. Events such as detecting a lead vehicle's brake light or a child on the roadside are considered less expected on the expectancy continuum [71], compared to standard signal detection tasks prescribed by ISO standard ISO/DIS 17488 [68]. However, this attempt is based on a single on-road study and has shown promising sensitivity for evaluating HUDs and their impact on safety. As we move towards a safety-centric evaluation framework, we believe this detection approach can be improved in several ways.

**First**, other driving assessments methodologies such as the NTSHA eye glance testing using a driving simulator [104], the visual occlusion technique [105], the lane change test [106], and the detection response task [68] have standardized guidelines on when and how to use these evaluation techniques. Similarly, our approach, based on detecting relevant driving-related events, would benefit from such standardization. Establishing guidelines related to experimental design, types of detection events, data analysis, and interpretation of results would aid researchers evaluating HUDs in both research and practice. This standardization would also facilitate cross-experiment comparisons of results.

**Second**, an essential factor with significant consequences for a safety-centric evaluation framework for HUDs is the variety and quantity of stimuli used in the detection approach. Unanticipated or surprise events are more realistic as they replicate real-world scenarios where accidents happen [107]. However, there is a balance to be struck between maintaining the element of surprise and the number of data points that can be collected. In user research, it is standard practice to collect multiple data points per participant to ensure the reliability of the data and to support robust statistical analysis. However, if too many events occur within a single trial or if trials are too similar, the events may become predictable, even if they differ from each other. This predictability could shift the approach from detection to vigilance. Therefore, future studies need to explore how expectancy can be quantified in this context and identify factors influencing expectancy, such as different stimulus types and their relationship to various driving scenarios.

Moreover, an essential aspect of improvement involves establishing clear criteria for interpreting results from event-detection approaches. Identifying significant differences across variable levels alone may not effectively inform safety-prioritized decision-making. There is a possibility that none of the interface design options under evaluation would meet a minimum safety standard when considering drivers' ability to detect critical elements in the real world while using the HUD. Therefore, rather than focusing solely on statistical measures like expected values or means (ANOVAs, t-tests, etc.), it is crucial to define clear thresholds for unsafe behaviors that may not align with traditional data analyses practices. These thresholds should guide decision-making for interface improvements, ensuring that safety remains the top priority. This consideration is crucial because, according to the psychology of surprise [107], accidents often involve the tail of the distribution: the least prepared, least skilled drivers, rather than the average ones. In very safe systems, unsafe behavior is not typically at the mean but rather in the extremes. For example, when analyzing response times to a lead car suddenly braking, the variability and particularly the extreme (long) values are likely more indicative of safety concerns than the mean values. These extremes of inattention and delayed responses are most likely to lead to accidents.

Lastly, our research evaluated detection using traditional 'detection performance' metrics, as well as 'inattentional blindness' and 'attention capture' with the aid of eye-tracking metrics. As we integrate more AR graphics into the driving environment, eye tracking becomes invaluable in understanding whether certain events were missed because participants were distracted by other interface elements or real road distractions, even if they were looking in the right direction. This distinction would help researchers better understand the impact of interface elements (such as color, contrast, and luminance, etc.) on the incidence of inattentional blindness in dynamic environments, which remains an uncharted territory.

## 7. LIMITATIONS

This study was conducted on a controlled test track, where drivers were instructed to perform detection tasks. Their behavior during the task may not accurately represent their performance when driving their own vehicle on public roads, where real-life stressors and traffic conditions can influence their actions.

Another potential limitation of our results is the possible novelty of the HUD. It is likely that this study was the first time some participants were exposed to a HUD. Since novelty was not the focus of the study, the number of inattentional blindness and attention capture instances observed during driving could differ if drivers were more familiar with the HUD. In fact, past research has shown that effective training and experience are crucial in countering the "novelty effect," which has been shown to reduce users' susceptibility to cognitive capture as they become more experienced with the head-up display [108].

In relation to the organization of the study, including the presentation of AR tasks and detection tasks in both central and peripheral fields of view, proved challenging to coordinate. Experimenters had to manage the task start, brake light activation, and proper location and orientation of peripheral targets. Due to occasional misalignments, some trials had to be discarded or readjusted during the study's execution.

While we utilized eye glance analysis to determine where drivers were fixating during events of interest, we could not ascertain the focal distances of fixation points to indicate whether drivers were focusing on the display or the external environment. Future research should investigate focal distance and its relationship to inattentional blindness, an aspect that, to the best of our knowledge, has not been explored thus far.